%% file: paper.tex
\documentclass{article}

\input{common}

\title{Seismic monitoring of \CO2 plume dynamics using ensemble Kalman filtering}
\date{\today}

\begin{document}

\maketitle

\begin{abstract}
Monitoring carbon dioxide (\CO2) injected and stored in subsurface reservoirs is critical for avoiding failure scenarios
and enables real-time optimization of \CO2 injection rates.
Sequential Bayesian data assimilation (DA) is a statistical method for combining information over time from multiple sources to estimate a hidden state, such as the spread of the subsurface \CO2 plume.
An example of scalable and efficient sequential Bayesian DA is the ensemble Kalman filter (\enkf/).
We improve upon existing DA literature in the seismic-\CO2 monitoring domain by applying this scalable DA algorithm to a high-dimensional \CO2 reservoir using two-phase flow dynamics and time-lapse full waveform seismic data with a realistic surface-seismic survey design.
We show more accurate estimates of the \CO2 saturation field using the \enkf/ compared to using either the seismic data or the fluid physics alone.
Furthermore, we test a range of values for the \enkf/ hyperparameters and give guidance on their selection for seismic \CO2 reservoir monitoring.
\end{abstract}

This work was supported by the National Science Foundation under Grant 2203821.

\section{Introduction}

Carbon capture and storage is a recently sought-after technology involving capturing carbon dioxide (\CO2) for long-term storage.
Geologic storage sites such as saline aquifers and depleted oil fields can store large amounts of \CO2 due to their size and the naturally high pressure deep underground \cite{solomon_co2_2014,raza_integrity_2016,raza_significant_2019}.
Monitoring the state of such \CO2 reservoirs is critical to avoid failure scenarios, such as leaks or man-made seismic activity (earthquakes) \cite{raleigh_experiment_1976,he_risk_2011}.
Monitoring also enables real-time optimization of reservoir performance using computer simulations.
However, due to the limited accessibility of underground reservoirs, traditional measurements of the reservoir state are confined to the surface and a relatively small number of boreholes compared to the size of the reservoir.
This limits the predictive power of simulations that are based on traditional measurements.
Since injected \CO2 replaces brine in saline aquifers, it affects the wave propagation in the rock, and the data from seismic waves reveal information at otherwise inaccessible points.
Thus, seismic measurements provide a non-intrusive method for determining the spread of the \CO2 in saline aquifers, providing higher resolution estimates than other non-intrusive methods such as measurements of the electromagnetic or gravitational field \cite{stephenson_highresolution_1993,gao_joint_2012,abubakar_inversion_2009,um_strategy_2014,filina_integration_2015}.

This paper combines fluid-flow simulations with time-lapse full waveform seismic data to predict \CO2 saturations in underground reservoirs as \CO2 is injected.
While most existing approaches use time-lapse seismic data by itself, the combination of physical simulation and
full waveform seismic data has recently begun to be studied \cite{huang_towards_2020,grana_prediction_2021}.
Challenges that hinder progress in this area include the cost of the high-resolution physical simulations and the cost of storing the covariance matrix necessary for properly combining information from the simulation and from system measurements.
These problems have been tackled in other domains, such as weather forecasting, using algorithms such as ensemble Kalman filters, particle filters, and variational optimization. These methods are all forms of Bayesian data assimilation.

Bayesian sequential data assimilation aims to estimate a state vector over time with quantified uncertainty given a sequence of noisy time-lapse measurements.
For our purposes, the state is the \CO2 saturation throughout the reservoir, and the measurements come from seismic waveform data.
The state follows known two-phase flow dynamics with uncertainty coming from the unknown subsurface permeability field, which describes the rate at which the \CO2 and brine flow through the porous rock.
To appropriately combine noisy time-lapse measurements with uncertain simulation predictions, we apply Bayesian sequential data assimilation.
Specifically, we demonstrate the computational efficiency and the predictive efficacy of the ensemble Kalman filter (\enkf/) technique on \CO2 reservoirs with multiple time-lapse surveyed seismic data.

The main issue with data assimilation for \CO2 reservoirs is that the sheer size of the problem makes the simulation and the data assimilation computations expensive. \CO2 reservoirs are represented in simulation by discretizing the reservoir into a three-dimensional grid and associating at the very least one \CO2 saturation value and one pressure value at each point in the grid. The system is governed by nonlinear dynamics, and thus demands a relatively high-resolution grid in order to be acceptably accurate. For example, a ``small" reservoir can have a volume of $10^{10}$ cubic meters (equivalent to a few dozen cubic kilometers) and be broken into cells on the order of $10^2 \sim 10^3$ cubic meters. The state vector (the \CO2 saturation) has $10^7$ degrees of freedom. Seismic measurements are collected by recording seismic waves with a resolution of about $10^{-3}$ seconds (equivalent to a few milliseconds) for a duration on the order of a second. A ``small" seismic survey records these waves for 10--100 sources for each of 100--1000 receivers, making the total length of the observation vector approximately $10^8$. While this is not prohibitively large for a single vector (a few hundred megabytes), data assimilation requires storing data many times that size, which can quickly cause issues with computation.
In our experiments, we model a two-dimensional (2D) reservoir with volume $\sim 10^8$ cubic meters, so the state vector is of size $\sim 10^5$, and the observation vector is of size $\sim 10^6$.
For classical data assimilation techniques that require storing the square of the data size, this 2D system requires almost one hundred gigabytes of computer memory, while the three-dimensional $10^{10}$ cubic meter system requires more than one terabyte of memory.

A classical data assimilation technique is the \defacronym{Kalman filter}{KF}, which computes the optimal state for a given set of measurements assuming linear transition and measurement functions with Gaussian noise.
With $\xs$ as the length of the system state vector and $\ys$ as the length of the measurement vector, the standard KF computes the Gaussian distribution of the state vector $x$ in $O(\xs^2 + \ys^3)$ time and $O(\xs^2 + \ys^2)$ storage, excluding the cost of the simulations.
Other data assimilation research on \CO2 reservoirs has handled the challenge of large covariance matrix sizes from seismic waveform observations by using a hierarchical sparse matrix structure \cite{li_kalman_2014,huang_towards_2020}
or using ensemble-based methods \cite{dupuy_bayesian_2021} with the neighborhood algorithm \cite{sambridge_geophysical_1999}.
However, these past data assimilation approaches with full waveform measurements of \CO2 reservoirs have ignored the \CO2 dynamics.

\CO2 reservoirs, like many real-world systems, are governed by nonlinear dynamics, which breaks the Gaussianity assumption of the KF. The KF can be successfully applied to nonlinear systems (in which case it is called the \emph{extended} KF) if the time between measurements is so small that the transition is approximately linear. But the high cost of seismic surveys leads to a long time between reservoir measurements, which breaks the linear approximation.
Furthermore, seismic observation is a nonlinear operator, further breaking the assumptions of the KF.

Weather forecasting is the well-researched domain that routinely handles nonlinearity, noise, and large data sizes for data assimilation.
Major weather forecasting centers rely on various data assimilation algorithms, including variational methods such as \defacronym{four-dimensional variational assimilation}{\fdvar/} and ensemble methods such as the \defacronym{ensemble Kalman filter}{\enkf/}.

Variational methods, on the one hand, minimize the observation error regularized with the distance from the predicted state. These methods can have a high implementation cost because they require adjoint operators for computing gradients. Additionally, for large systems, storing the state covariance matrix becomes infeasible.
\textcite{bannister_review_2008-1, bannister_review_2008} describes how weather forecasting models reduce the matrix storage size by diagonalizing the covariance matrix using a physics-based transform.
However, this transform becomes less accurate as model resolutions increase, and as a result, many weather systems have switched to an ensemble-based estimate of the covariance.
Furthermore, the variational methods do not estimate the posterior uncertainty, so they must be combined with ensemble methods to obtain a covariance estimate.

Ensemble methods, on the other hand, are comparatively much easier to implement because they do not require adjoint operators.
The basic \enkf/ updates a sample of system states forward in time and uses the sample covariance to update the samples based on observations. Although this algorithm is optimal only for linear models, it does not rely on linearizing the transition or observation models and thus handles nonlinearities much better than the standard KF. Furthermore, using a low-rank form for the covariance simplifies the update cost to $O(\xs\es)$ for an ensemble of size $\es$.
The \enkf/ has been shown to scale well with $\xs$ both in accuracy and computational cost for many data assimilation problems \cite{evensen_ensemble_2003, butala_tomographic_2009, awasthi_use_2021}.
Several variants of the \enkf/ exist, such as those described by \textcite{evensen_sampling_2004}, but we leave other variants to future work.

In seismic measurement domains, several variations on the KF of varying complexity have been used to address nonlinearity, noise, and large data, but none yet have addressed the combination of large \CO2 reservoirs with \CO2 dynamics and full waveform seismic measurements.
\textcite{eikrem_iterated_2019} address the issue of observation nonlinearity by using the iterated extended KF to estimate wave velocities from the full waveform data. This method linearizes the model at multiple points, thereby better approximating the nonlinearity, but they note it requires an adjoint observation model and the matrix storage causes difficulty for scaling to large data.
\textcite{li_compressed_2015} address the issue of scalability with a low-rank KF to estimate \CO2 reservoir state. They show its performance on seismic travel time measurements or 115 wells on a small 0.2 square kilometers domain, but this method has not been shown on high-dimensional seismic data.
\textcite{alfonzo_seismic_2020} apply the ensemble transform KF to estimate porosity based on acoustic impedance inverted from seismic measurements, but this does not estimate the dynamic state of the reservoir.
\textcite{guzman_coupled_2014,ma_dynamic_2019,grana_prediction_2021} apply the \enkf/ to estimate \CO2 reservoir state and geological parameters using \CO2 dynamics but not full waveform observations.
A scalable application of the KF using full waveform data and \CO2 dynamics has yet to be achieved.

In the machine-learning (ML) domain, research that uses \CO2 dynamics has focused on building surrogate flow models for use in Bayesian inversion algorithms to estimate static properties, such as porosity and permeability or some other parameter for the surrogate model \cite{tang_deep-learning-based_2022, tang_deep-learning-based_2020, seabra_ai_2024, tang_deep_2021,tang_deep_2022, yin_learned_2022,liu_joint_2023,sakai_co2_2024}.
The surrogate models allow the use of inversion algorithms that are more expressive but more expensive than the KF,
for instance, rejection sampling \cite{tang_deep-learning-based_2022}; randomized maximum likelihood \cite{tang_deep-learning-based_2020, seabra_ai_2024}; ensemble smoothing with multiple data assimilation (ES-MDA) \cite{seabra_ai_2024,tang_deep_2022,tang_deep_2021}; and variational methods \cite{yin_learned_2022, liu_joint_2023,sakai_co2_2024}.
While one reference uses full-waveform seismic data \cite{yin_learned_2022}, most do not, instead relying on less informative data such as well data \cite{tang_deep-learning-based_2020, seabra_ai_2024,sakai_co2_2024} or surface deformation data \cite{tang_deep_2022,tang_deep-learning-based_2022}, while those using seismic data approximate the seismic image directly \cite{tang_deep_2021,liu_joint_2023}.
We are also investigating ML-based DA research using full-waveform seismic data and \CO2 fluid dynamics; see \cite{gahlot_inference_2023,gahlot_digital_2024} for preliminary results.
While ML-based methods show promise, they have not reached the level of robustness, scalability, and interpretability that the \enkf/ has has shown in the field of weather forecasting.

In this research, we show that the \enkf/ with full waveform data and \CO2 dynamics estimates the \CO2 reservoir state with higher accuracy than two non-data-assimilation baselines without needing a prohibitively large ensemble. This study contributes to the broader objective of incorporating full waveform data and known physics into the scalable and computationally efficient \enkf/.
The paper is structured as follows: Section 2 describes the mathematical background for filters, two-phase flow, and seismic measurements; Section 3 describes how we structure the two-phase flow and seismic operators in the \enkf/; Section 4 describes our high-dimensional synthetic experiment; Section 5 describes the results of our tests with the \enkf/; and Section 6 concludes.

\section{Background}

\subsection{Filters}

Data assimilation is the process of estimating a hidden system state using available observations as well as knowledge of the system's dynamics. The simulated dynamics are represented with the Markovian transition function $f$, which updates the state $x^n$ at time step $n$ to the next time step. System measurement is represented with the observation function $h$. Both functions may depend on noise $\eta$ sampled at each time step independent of $x$, which represents any stochasticity of the system or unknown system dynamics.
The transition and observation operators may depend on time, but we will typically omit the time parameters as they are clear in context. We also leave off the noise parameter as necessary when indicating a non-noisy evaluation.
Mathematically, the operators are written as
\begin{align}
x^{n+1} &= f(x^n, \eta_f; t_n, t_{n+1}) \text{ and } \label{eq:trans_f}\\
y^{n+1} &= h(x^{n+1}, \eta_h; t_{n+1}), \label{eq:obs_h}
\end{align}
with $\eta_f$ and $\eta_h$ being noise samples for $f$ and $h$.

Using a sequence of observation data vectors $\{y^1, y^2, \ldots\,  y^n\}$, data assimilation algorithms estimate the hidden states $\{x^1, x^2, \ldots, x^n\}$.
Typically, we are unable to observe the full hidden state, making this an ill-posed inverse problem requiring regularization. Regularization can be expressed as a prior distribution in Bayes' formula, which states
\begin{align}
p(x | y) &= \frac{ p(y | x) p(x)}{p(y)}.
\end{align}
The prior $p(x)$ can be chosen based on physics or smoothness knowledge.
In sequential data assimilation, it is conditioned on previous data and the prediction of a simulation. Let $y^{1:n}$ denote all observations up to time step $n$. Given a distribution $p(x^{n-1} | y^{1:n-1})$, the prior, or predictive, distribution $p(x^n | y^{1:n-1})$ conditioned on all data from previous time steps is given by
\begin{align}
p(x^n | y^{1:n-1}) &=  \int p(x^n | x^{n-1}) p(x^{n-1} | y^{1:n-1}) \;dx^{n-1}, \label{eq:complex_predict}
\end{align}
where $p(x^n | x^{n-1})$ is the probability density determined by the transition function from \cref{eq:trans_f}.
With that prior, Bayes' formula for sequential data assimilation becomes
\begin{align}
p(x^n | y^{1:n}) &= \frac{ p(y^n | x^n) p(x^n | y^{1:n-1})}{p(y^n|y^{1:n-1})}, \label{eq:complex_update}
\end{align}
with $p(x^0)$ fixed to represent the initial knowledge of the system before any measurements.

Since we always incorporate available measurements, we simplify the notation by defining the distributions at time step $n$ to be implicitly conditioned on measurements from the previous time steps. 
We define $p(x^n) \equiv p(x^n | y^{1:n-1})$ and $p(x^n | y^n) \equiv p(x^n | y^{1:n})$.
That also allows us to drop the time step superscript when looking at data at a single time step.
The form of data assimilation in \cref{eq:complex_predict,eq:complex_update} can be written as two repeated phases, shown both in \cref{alg:generic_da} and here mathematically as
\begin{align}
\text{predict:}&\quad  p(x^n) = \int p(x^n | x^{n-1}) p(x^{n-1}|y^{n-1}) \;dx^{n-1}, \label{eq:predict}\\
\text{update:}&\quad p(x^n | y^n) \equiv p(x | y) = \frac{ p(y | x) p(x)}{p(y)},
\label{eq:update}
\end{align}
with implicit $n$ superscripts in the update phase.
The predict phase advances the distribution $p(x^{n-1}|y^{n-1})$ to $p(x^n)$ using the transition function from \cref{eq:trans_f}.
The update phase updates the prior $p(x^n)$ to the posterior $p(x^n | y^n)$ using the likelihood $p(y^n|x^n)$, which is based on the observation function from \cref{eq:obs_h}.
In \cref{sec:kf}, we explain the update phase for the Kalman filter and the \enkf/.
In \cref{sec:trans,sec:obs}, we introduce the transition and observation operators we use with the \enkf/.

\begin{algorithm}[t]
\Input{\Prior, data $y^n$ at time $t_n$}
\BlankLine
\Posterior $\gets$ \Prior \;
$t_{n-1} \gets$ 0\;
\For{($y^n$, $t_n$) \KwForIn \NewData}{
    \Comment{\Cref{eq:predict}}
    \Prediction $\gets$ \Advance{\Posterior, $t_{n-1}$, $t_n$}\;
    \Comment{\Cref{eq:update}}
    \Posterior $\gets$ \Assimilate{\Prediction, $y^n$}\;
    $t_{n-1} \gets t_n$ \;
}
\caption{Generic data assimilation loop}\label{alg:generic_da}
\end{algorithm}

\subsection{Kalman filters}\label{sec:kf}
The standard \defacronym{Kalman filter}{KF} \cite{kalman_new_1960} is a sequential application of Bayes' rule that assumes each distribution is Gaussian in the state $x$. Consider linear transition and observation functions with Gaussian noise, written as 
\begin{align}
x^{n+1} &= f(x^n, \eta_f) = Fx^n + \eta_f \; \text{ and } \label{eq:trans_F}\\
y^{n+1} &= h(x^{n+1}, \eta_h) = Hx^{n+1} + \eta_h, \label{eq:trans_H}
\end{align}
where $F$ and $H$ are linear operators; $Q$ and $R$ are the known noise covariances; $\mathcal{N}(\mu, B)$ denotes a multivariate Gaussian distribution with mean $\mu$ and covariance $B$; $\eta_f \sim \mathcal{N}(0, Q)$; and $\eta_h \sim \mathcal{N}(0, R)$.

In this linear case, an input Gaussian distribution $p(x^{n-1} | y^{n-1})$ always transforms to a Gaussian prior $p(x^n)$ and posterior $p(x^n | y^n)$. If we parametrize the distribution with $\mu$ and $B$ as $p(x) = \mathcal{N}(\mu, B)$, then we have linear solutions for \cref{eq:predict,eq:update}.
For the linear transition function of \cref{eq:trans_F}
where the distribution at the previous time step $n-1$ is known as $p(x^{n-1} | y^{n-1}) = \mathcal{N}(\mu^{n-1}_a, B^{n-1}_a)$, then the predicted (or \emph{forecasted}), distribution is $p(x^n) = \mathcal{N}(\mu^n_f, B^n_f)$, where
\begin{align}
\mu^n_f &= f(\mu^{n-1}_a) \; \text{ and }
B^n_f = F B^{n-1}_a F^T + Q, \label{eq:transitionmuB}
\end{align}

In the update phase, the KF computes the posterior $p(x^n|y^n)$ given the prior $p(x^n)$, the likelihood $p(y^n|x^n)$, and the observation $y^n$. For the linear observation function of \cref{eq:trans_H}, the likelihood is $p(y|x) = \mathcal{N}(Hx, R)$. Let $\mu_0 = \mu^n_f$ and $B_0=B^n_f$ be the forecasted mean and covariance of $x_0 = x^n$ from the prediction step. Then the posterior distribution is $p(x|y) = \mathcal{N}(\mu, B)$, where the moments can be calculated as
\begin{align}
\mu &= \mu_0 + K(y - h(\mu_0)) \text{ and }  \label{eq:kalmanmu} \\ \label{eq:kalmanB}
B &= (I - KH) B_0, 
\end{align}
with $K$ known as the Kalman gain matrix,
\begin{align}
K &= B_0 H^T (H B_0 H^T + R)^{-1}. \label{eq:kalmanK} 
\end{align}

The Kalman gain matrix can be expressed in terms of the covariance of the predicted $x_0 \sim p(x_0)$ and $y_0 \sim p(y|x_0)$. Recall the definition of the covariance $\cov(\cdot)$ and the cross-covariance $\cov(\cdot, \cdot)$ as the expectation of the outer product of the deviations from the mean,
\begin{align}
B_0 = \cov(x_0) = \Expect_{x_0 \sim p(x_0)} (x_0 - \mu_0)(x_0 - \mu_0)^T.
\end{align}
Then $B_0H^T$ is the cross-covariance $\cov(x_0, y_0)$, and $H B_0 H^T + R$ is the forecast observation covariance $\cov(y_0)$. Thus, $K$ can be expressed as
\begin{align}
K &= \cov(x_0, y_0) \cov(y_0)^{-1}. \label{eq:kalmanK_cov} 
\end{align}
This is equivalent to \cref{eq:kalmanK} in the Gaussian case, but it can be easily extended to non-Gaussians, for instance, in the \enkf/.

There are several KF variants based on \cref{eq:transitionmuB,eq:kalmanmu,eq:kalmanB,eq:kalmanK} made to deal with invalid assumptions in the standard KF. Specifically, they are made to handle nonlinearity (and therefore, non-Gaussianity) and the inability to store the full covariance matrix.
Below, we explain three methods to handle nonlinearity to show how the \enkf/ method uses an ensemble to avoid storing large matrices.

The \defacronym{extended Kalman filter}{EKF} (first described by \textcite{smith_application_1962}) extends \cref{eq:kalmanmu,eq:kalmanB,eq:kalmanK,eq:transitionmuB} to nonlinear transitions and observations by linearizing the functions about the prior mean. $H$ is replaced by the Jacobian $\frac{dh}{dx}$ evaluated at $x = \mu_0$ and similarly for $F = \frac{df}{dx}$.
The pseudocode for the predict and update phases is shown in \cref{alg:kf}.
Since EKF linearizes the transition and observation operators, it fails when the system exhibits significant nonlinear behavior when transitioning between observation time steps or between the prior mean and the posterior mean.
Similarly to the original KF, the EKF can also fail when the noise statistics are specified incorrectly, as described by \textcite{valappil_systematic_2000}.

The \defacronym{iterated extended Kalman filter}{IEKF} extends the EKF by taking smaller steps in time and in posterior updates. It repeats 
\cref{eq:transitionmuB} a fixed number of times with a smaller time step size, correspondingly smaller $Q$, and a recomputed linearization $F$ at each iteration.
Similarly, \cref{eq:kalmanmu,eq:kalmanB,eq:kalmanK} are iterated a fixed number of times with a scaled-up observation error covariance $R$ and a recomputed linearization $H$ at each iteration's computed mean.
This can be computationally expensive, but reduces linearization errors for nonlinear problems compared to the EKF.

\begin{algorithm}[t]
\Input{forward physics operator $f(x; t_0, t)$ that simulates $x$ from time $t_0$ to time $t$ with noise covariance matrix $Q$, measurement operator $h(x)$ with noise covariance matrix $R$}
\BlankLine
\Fn{\Advance{$(\mu_0, B_0)$, $t_0$, $t$}} {
\Comment{\Cref{eq:transitionmuB}}
$\mu \gets f(\mu_0; t_0, t)$\;
$F \gets \frac{df}{dx}(\mu_0)$\;
$B \gets F B_0 F^T + Q$\;
\Return $(\mu, B)$
}
\BlankLine
\Fn{\Assimilate{$(\mu_0, B_0)$, $y$}} {
$H \gets \frac{dh}{dx}(\mu_0)$ \;
\Comment{\Cref{eq:kalmanK}}
$K \gets B_0 H^T(H B_0 H^T + R)^{-1}$\;
\Comment{\Cref{eq:kalmanmu}}
$\mu \gets \mu_0 + K(y - h(\mu_0))$\;
\Comment{\Cref{eq:kalmanB}}
$B \gets (I - KH)B_0(I - KH)^T + KRK^T$\;
\Return $(\mu, B)$\;
}
\caption{Extended Kalman filter implementation for use with \cref{alg:generic_da}}\label{alg:kf}
\end{algorithm}

The \defacronym{ensemble Kalman filter}{\enkf/} \cite{evensen_sequential_1994,burgers_analysis_1998} handles non-linearity by using multiple values of the operators at different points, which can be contrasted with the EKF strategy of using the Jacobian at the mean.
The prior is represented by an ensemble of samples. The predict phase advances each sample forward in time using the transition operator.
The update phase simulates observations of each sample and applies a linear Bayesian update to each sample to better match the true observations.
The update assumes the state and observation samples are jointly Gaussian, so this method is optimal only for the linear KF case.
However, because the samples do not have to be Gaussian, this method has the capability of expressing non-Gaussian distributions.
Specifically, if the transition operator is nonlinear, the prior $p(x)$ for the update step is not Gaussian, and if $h$ is nonlinear, the likelihood $p(y|x)$ for the update step is not Gaussian.

The Kalman update in \cref{eq:kalmanmu,eq:kalmanB,eq:kalmanK} is equivalent to taking the mean of a linear transform $z$ in the form $z(x, y) = x + K(y - Hx - \eta_h) = (I - KH) x + K(y - \eta_h)$. If $x$ is sampled from $\mathcal{N}(\mu_0, B_0)$, the resulting $z$ is distributed according to the posterior $\mathcal{N}(\mu, B)$. Thus, we have a method of computing posterior samples based on prior samples and measurements.

Let the ensemble be forecasted samples $x_{f,i}$ indexed by $i$ from 1 to $\es$.
They can be described as samples of $\mathcal{N}(\mu_0, B_0)$ with forecasted sample mean $\mu_0 = \sum x_{f,i}/\es$ and sample covariance $B_0 = X_fX_f^T$, where
$X_f$ is a matrix where the $i$-th column is $(x_{f,i} - \mu_f)/\sqrt{\es-1}$.
\Cref{eq:kalmanmu} applies to each sample to update the samples with the observation $y$, written as
\begin{align}
x_{a,i} &= x_{f,i} + K(y - y_{f,i}) \text{ for } i \in \{1,2,\ldots,\es\}, \label{eq:enkf_update}
\end{align}
where $y_{f,i} = h(x_{f,i}, \eta_{h,i})$ is a simulated observation, $\eta_{h,i}$ is a sample of the noise for the observation operator, and $x_{a,i}$ is a sample of the posterior.
Note that the observation operator here on each sample is simulated with noise, whereas in \cref{eq:kalmanmu}, the mean is observed without noise, since the noise is assumed to have zero mean.
The EKF uses \cref{eq:kalmanK} to compute the Kalman matrix $K$ with a linearized form of $h$. Directly plugging in $B_0 = X_f X_f^T$ yields
\begin{align}
K &= X_f (HX_f)^T ( (HX_f)(HX_f)^T + R)^{-1}.
\end{align}
For the EKF, $H$ is the Jacobian of $h$ at $x = \mu_0$, and it is used to estimate how differences in the value of $x$ correspond to differences in $y$. However, for nonlinear functions, this may poorly describe the behavior of $h$ in a larger neighborhood around $\mu_0$.
Instead, the \enkf/ better explores the nonlinearity of the function by sampling those differences at various points, using the approximation $H(x_{f,i} - \mu_0) \approx y_{f, i} - \mu_y$, where $\mu_y = \sum_j y_{f, j}/\es$ is the mean of the predicted observations.
This approximation yields the version of the Kalman matrix shown in \cref{eq:kalmanK_cov}.
Let $B_y = Y_f Y_f^T$ be the sample observation covariance,
where $Y_f$ is a matrix where the $i$-th column is $(y_{f,i} - \mu_y)/\sqrt{\es-1}$.
The covariance form of the Kalman matrix removes possibly costly applications of $H$ by replacing the linear approximation $HX_f$ with the difference matrix $Y_f$, yielding
\begin{align}
K &= X_f Y_f^T (Y_fY_f^T)^{-1}.
\end{align}
Note that if the ensemble size is smaller than the observation size, the sample observation covariance is guaranteed to be singular.
If the (non-singular) observation noise covariance $R$ is known, the sample observation covariance can be computed as $\hat Y_f \hat Y_f^T + R$, where $\hat Y_f \hat Y_f^T$ is the non-noisy sample observation covariance obtained from simulating observations without noise.
Or, a non-singular form of the sample observation covariance can be obtained by adding a small regularization as $Y_fY_f^T + \epsilon_h R$,
where $Y_f Y_f^T$ is the noisy sample observation covariance and $\epsilon_h$ is some small constant.

Pseudocode for the \enkf/ predict and updates phases is shown in \cref{alg:enkf}.

The Kalman filter update in \cref{eq:kalmanmu} can be thought of as solving the linearized optimization problem
\begin{equation}
\mu = \arg \min_x \, \|h(\mu_0) + H(x - \mu_0) - y\|_{R^{-1}}^2 + \|x - \mu_0\|_{B_0^{-1}}^2,
\end{equation}
and then the covariance estimate $B$ is the inverse Hessian of the optimization objective at $\mu$.
Similarly, the \enkf/ can be thought of as updating each ensemble member by solving the linearized optimization problem
\begin{align}
c_i &= \argmin_{c \in \mathbb{R}^{\es}} \, \|\mu_y + Y_f c - y\|_{R^{-1}}^2 +  \| \mu_0 + X_fc - x_{f,i}\|_{B_0^{-1}}^2,  \label{eq:enkfopt}
\end{align}
with $x_{a,i}=\mu_0+X_fc_i$.
Compared to the extended Kalman filter, the \enkf/ must do multiple linear solves to apply the Kalman gain matrix to each ensemble member, but the \enkf/ gains the benefit of the covariance matrix being implicitly represented and implicitly updated.

Although the \enkf/ efficiently represents the covariance, the \enkf/ encounters similar challenges as other variants of the KF, specifically in choosing noise statistics.
When noise statistics are underestimated, the filter's estimated state covariance becomes increasingly small, such that new observations have negligible effect on the estimated state and the filter diverges from the truth.
This is especially an issue for the \enkf/ with small sample size.
Note that the update to each $x$ resides in the range of $X$, so it is important that $X$ have full column rank, requiring the ensemble members to stay separated.

To manage this filter divergence, a conventional approach is to increase the variance in the state.
Covariance inflation, described by \textcite{anderson_monte_1999}, artificially inflates the covariance of the ensemble by an ad hoc factor.
\textcite{hamill_hybrid_2000,whitaker_ensemble_2002,whitaker_reanalysis_2004} showed that this simple, efficient approach improved the predictive accuracy of atmospheric data assimilation models.
In the context of seismic measurements, \textcite{alfonzo_seismic_2020} increased the state covariance by estimating observation noise using observation error, thereby avoiding filter divergence.

Filter divergence is especially a problem when observations are assimilated often.
For seismic surveys, we assume observations are expensive and therefore temporally sparse, and we do not assimilate enough observations to encounter filter divergence.
Our \enkf/ simulates reservoir dynamics for a year with unknown permeability and then assimilates a seismic survey.
The uncertainty in the parameters governing the reservoir dynamics over this large of a time period is enough to ensure the filter covariance does not become too small. The next two subsections describe the reservoir dynamics and the seismic observation operator we use for the \enkf/.

\begin{algorithm}[t]
\Input{forward physics operator $f(x, \eta_f; t_0, t)$ that simulates $x$ from time $t_0$ to time $t$ with noise distribution $p(\eta_f)$, measurement operator $h(x, \eta_h)$ with noise distribution $p(\eta_h)$ with estimated covariance $R$}
\BlankLine
\Fn{\Advance{$\{x_1, x_2, \ldots, x_{\es}\}$, $t_0$, $t$}} {
\For{$i$ \KwForIn $\{1, 2, \ldots, \es\}$} {
Sample $\eta_f \sim p(\eta_f)$\;
$x_i \gets f(x_i, \eta_f; t_0, t)$\;
}
\Return $\{x_1, x_2, \ldots, x_{\es}\}$\;
}
\BlankLine
\Fn{\Assimilate{$\{x_1, x_2, \ldots, x_{\es}\}$, $y$}} {
\For{$i$ \KwForIn $\{1, 2, \ldots, \es\}$} {
Sample $\eta_h \sim p(\eta_h)$\;
$y_i \gets h(x_i, \eta_h)$\;
}
$\mu_x \gets \sum_{i=1}^{\es}x_i / \es $\;
$\mu_y \gets \sum_{i=1}^{\es}y_i / \es $\;
\For{$i$ \KwForIn $\{1, 2, \ldots, \es\}$} {
$X_f$[:, i] $\gets (x_i - \mu_x) / \sqrt{\es - 1}$\;
$Y_f$[:, i] $\gets (y_i - \mu_y) / \sqrt{\es - 1}$\;
}
$K \gets X_f Y_f^T(Y_f Y_f^T + R)^{-1}$\;
\For{$i$ \KwForIn $\{1, 2, \ldots, \es\}$} {
$x_i \gets x_i + K(y - y_i)$\;
}
\Return $\{x_1, x_2, \ldots, x_{\es}\}$\;
}
\caption{Ensemble Kalman filter implementation for use with \cref{alg:generic_da}}\label{alg:enkf}
\end{algorithm}

\subsection{Two-phase flow}\label{sec:trans}

Multi-phase porous flow is a standard approximation for \CO2 reservoir dynamics.
For our purposes, the reservoir consists of briny water and injected supercritical \CO2 flowing through porous rock.
Simulations may also account for aqueous \CO2, but in our experiments, we consider the two fluids to be immiscible.

For each fluid, a saturation field describes what proportion of the pore space is filled with that fluid. 
Due to differences in the fluids' adhesion to the rock, each fluid has its own pressure field, although in our experiments, we take the pressure fields to be identical.
Let $S_g$ and $P_g$ be the \CO2 saturation and pressure, usually written with no subscript because they are the main quantities of interest.
For two-phase flow, the water saturation $S_w$ and pressure $P_w$ can be directly computed as $S_w = 1 - S_g$ and $P_w = P_g + P_c$, where $P_c$ is the capillary pressure. 
Capillary pressure can be a complex function \cite{beliaev_theoretical_2001}, but we do not attempt to address that field of research. Instead, we focus on the effects of permeability and take capillary pressure to be 0.

The governing PDE is the mass balance equation with Darcy flow,
\begin{align}
\frac{\partial(\phi\rho_i S_i)}{\partial t} - \nabla \cdot (\rho_i v_i) &= \rho_i q_i,\\
v_i &= -  k_{ri}/\mu_i K \nabla (P_i - \rho_i g Z),\\
S_i, P_i&\text{ given at $t = 0$,}
\end{align}
with $g$ being gravitational acceleration and $Z$ being depth.
Fluid $i$ has fluid density $\rho_i$, velocity $v_i$, volume injection rate $q_i$, relative permeability $k_{ri}$, and fluid viscosity $\mu_i$.
The permeability field $K$ describes the relationship between fluid flow rate and an applied pressure gradient.
The flow rate is slowed by viscosity and relative permeability.
Permeability is usually expressed in millidarcies and is anisotropic, typically favoring horizontal directions due to sedimentary layering. We represent it with a diagonal tensor separated into vertical and horizontal components,
\begin{align}
K = \begin{bmatrix} K_v & 0 \\ 0 & K_h \end{bmatrix},
\end{align}
with $K_h$ being a heterogeneous field of the horizontal permeability and $K_v / K_h$ being a constant less than 1.
The porosity $\phi$ is a field with values in the range 0--1 representing how much of the rock is accessible by the fluids (typically about 20\%).
The relative permeability is a nonlinear function of the saturation.
In line with existing literature, we use a modified Brooks-Corey model with residual saturation $\alpha$ for both \CO2 and brine, given by
\begin{align}
k_{ri} &= \text{clamp}((1 - 2\alpha)^{-1}(S_i - \alpha),\; 0,\; 1)^2
\\&= \begin{cases}
    1 & \text{ if $S_i \ge 1 - \alpha$} \\
    0 & \text{ if $S_i \le \alpha$} \\
    (1 - 2\alpha)^{-2}(S_i - \alpha)^2 & \text{ else}
\end{cases}.
\end{align}
Let $\mathcal{M}$ solve the PDE from time $t_n$ to $t_{n+1}$ with initial conditions $S^n$ and $P^n$, permeability $K$, and porosity $\phi$, written as
\begin{align}
    S^{n+1}, P^{n+1} = \mathcal{M}(S^n, P^n, K, \phi; t_n, t_{n+1}). \label{eq:flow_M}
\end{align}
The resulting saturation is then used to simulate seismic measurements, as described in the next subsection.

\subsection{Seismic observation}\label{sec:obs}

Surface-seismic data allows measuring subsurface changes without drilling wells. 
In active-source surface seismic, a seismic source, such as an explosive or air gun, on the surface sends seismic waves into the reservoir.
Spatial differences in subsurface wave velocity cause the waves to be reflected or transmitted back to receivers on the surface.
The data recorded for this measurement consists of a time series for each possible source-receiver pair.

The wave dynamics are governed by the wave equation.
To simulate seismic measurements, the wave equation must be solved for each source to determine the data collected at the receiver locations.
In terms of the acoustic pressure field $\delta_p$, the governing wave dynamics are
\begin{align}
  \frac{m}{\rho} \frac{\partial^2 \delta_p}{\partial t^2} - \nabla \cdot \left(\frac{1}{\rho} \nabla \delta_p\right) + w \frac{\partial \delta_p}{\partial t} = \frac{q}{\rho},
\end{align}
where $m$ is the P-wave squared slowness field, $\rho$ is the density field, $w$ is a spatially-varying dampening parameter that is nonzero only in the absorbing boundary layer, and $q$ is an acoustic source.
The boundary layer is used to simulate an infinite domain, ensuring any waves that would escape the system of interest are absorbed by the boundary instead of being reflected back into the system.

Seismic measurements can detect \CO2 because the density and squared slowness at each point is dependent on the material composition at that point, including the rock composition and the relative concentrations of \CO2 and brine.
This relation is specified by a rock physics model, and a typical rock physics model for \CO2 reservoirs is the patchy-saturation model.
Past research has investigated the effects of pressure in the rock physics model \cite{macbeth_classification_2004,corte_bayesian_2023}, but we neglect that here to focus on the leading order effect.
The patchy-saturation rock physics model, described in Chapter 1 of \cite{avseth_quantitative_2010}, expresses the density and squared slowness as a pointwise function of the \CO2 saturation and a baseline model representing the density and squared slowness before injection.
This model is a nonlinear interpolation of the squared slowness with 0\% \CO2 and 100\% \CO2 in the pore space, based on the P-wave modulus of the rock and fluids.

The P-wave modulus $\lambda$ is related to the density and squared slowness by $\lambda = \rho / m$, and the shear wave modulus $G$ is related to the density and shear wave squared slowness $m_G$ by $G = \rho / m_G$. We use a typical relation of $m_G = 3m$, but in general, the relation can be more complex.
The bulk modulus $B$ is related to the P-wave and shear wave moduli by $B = \lambda - 4/3 G$.

Gassman's equation for porous media describes the relation between composite bulk moduli and constituent bulk moduli based on porosity \cite{gassmann_uber_1951}.
Here, the constituent bulk moduli are $B_0$ for the mineral making up the porous rock, $B_w$ for water, and $B_g$ for \CO2.
Given the density and velocity before injection, the bulk modulus $B_{wr} = \frac{5}{9}\rho/m$ of the pre-injection brine-rock system is computed using the relations $B_{wr} = \lambda_{wr} - 4/3 G$, $m_G = 3m$.
Then Gassman's equation below can be solved for the bulk modulus $B_{gr}$ of the rock fully saturated with \CO2,
\begin{align}
\frac{\phi^{-1} B_w}{B_0 - B_w}
- \frac{B_{wr}}{B_0 - B_{wr}}
&= 
\frac{\phi^{-1} B_g}{B_0 - B_g}
- \frac{B_{gr}}{B_0 - B_{gr}}.
\end{align}
Replacing brine with \CO2 does not change the shear modulus, so the resulting P-wave modulus of rock saturated with \CO2 is $\lambda_{gr} = B_{gr} + 4/3 G$.

The patchy-saturation model uses an arithmetic average for the density and a harmonic average for the P-wave modulus,
\begin{align}
\rho(S; \rho_{wr}) &= \rho_{wr} + S \phi (\rho_g - \rho_w), \\
\lambda(S; \lambda_{wr}) &= \left[(1 - S)\lambda_{wr}^{-1} + S \lambda_{gr}^{-1} \right]^{-1},
\end{align}
where $\lambda_{wr} = \rho_{wr} / m$ with pre-injection density $\rho_{wr}$ and squared slowness $m_{wr}$.
The harmonic average is a result of assuming the pressure is equalized in the mixed fluid.
Increasing the \CO2 saturation decreases the density and tends to decrease the P-wave modulus, depending on porosity.
The squared slowness is a quadratic function of the saturation,
\begin{equation}
\begin{split}
m&(S; \rho_{wr}, m_{wr}) = \rho(S)/\lambda(S)
\\ &= \big[\rho_{wr} + S \phi (\rho_g - \rho_w)\big]\big[(1 - S)\lambda_{wr}^{-1} + S \lambda_{gr}^{-1} \big].
\end{split}
\end{equation}

Given a squared slowness field $m$ from the rock physics model, let $d = \mathcal{H}(m, \rho) + \eta$ be the waveforms representing simulating the seismic waves and obtaining measurements at the receivers with noise $\eta$.
In terms of the \CO2 saturation $S$,
\begin{align}
    d(S; \eta) = \mathcal{H}(m(S), \rho(S)) + \eta, \label{eq:fullwaveform_H}
\end{align}
where $d$ being a vector with components $d_{ij}(t)$ as the acoustic pressure measurement at receiver $i$ from source $j$ at time $t$.
These are the full-waveform measurements we use for the \enkf/.

\section{\enkf/ applied to \CO2 plume with seismic observations}

We apply the scalable, efficient \enkf/ to the two-phase flow system with seismic observations and show that it gives good estimates of the \CO2 saturation in a high-dimensional (high-resolution) system with temporally sparse seismic measurements.
As our seismic model does not depend on the \CO2 pressure, we focus solely on estimating the \CO2 saturation over time.
We improve upon the existing literature by addressing all three issues of (1) relatively large problem size with (2) \CO2 dynamics and (3) seismic data.
Details on the transition and observation operators are below.

\subsection{Transition operator}

We use \cref{eq:flow_M} as the transition operator.
The permeability and porosity fields cannot be known exactly, which must be accounted for in our \enkf/.
We focus on the uncertainty created by the unknown permeability field, and we assume the porosity field is known to be a constant 25\%.
Accounting for the uncertainty in the permeability amounts to initializing each ensemble member with a different realization of the permeability field, with each ensemble member's permeability field fixed across time.

We assume that knowledge of the permeability is represented as a distribution from which we can sample permeability fields.
We use a probabilistic relation based on \cite{yin_time-lapse_2024} to generate permeability values from velocity values.
Specifically, we generate the permeability $K(\vec{r})$ at spatial coordinate $\vec{r}$ as a pointwise random function of a given velocity field $v(\vec{r})$ using
\begin{equation} \label{eq:VtoK}
K =
\begin{cases}
    10^{-2}  e^{v - 3.5} & \text{if $v < 3.6$} \\
    10^{-4 (3.85 - v)} c^{2 (v - 3.35)} & \text{if $3.6 \le v < 3.85$} \\
    c e^{v - 3.7 - w} & \text{if $v \ge 3.85$} \\
\end{cases}
\end{equation}
where $K$ is in millidarcies, $v$ is in km/s, and $c$ and $w$ are random fields representing random coefficients for the relation defined in \cite{yin_time-lapse_2024}.
Specifically, $c$ is a Gaussian random field, fixed across all ensemble samples, with mean 1200, standard deviation 3000, and a Gaussian covariance kernel with length scale 62.5 meters in the horizontal direction and 31.25 meters in the vertical direction.
The pointwise random field $w$ is generated from a discrete uniform distribution between 0 and 1.4 with step size of 0.1.
The ground-truth permeability is generated via the same methodology with the standard deviation of $c$ replaced with 5.

We assume to know all the transition function parameters except for the saturation, pressure, and permeability. Therefore, an ensemble member's state $x^n$ can be represented by a tuple $x^n = (S^n, P^n, K)$. The transition function $f$ updates the saturation and pressure given the permeability and leaves the permeability unchanged.
Using $\mathcal{M}$ described by \cref{eq:flow_M}, the transition operator can be written as
\begin{align}
    x^{n+1} = f(x^n) = (\mathcal{M}(S^n, P^n, K, \phi; t_n, t_{n+1}), K).
\end{align}
There is no stochasticity in this system, so we drop the $\eta_f$ term from \cref{eq:trans_f}.

\paragraph*{Discretization and solution}
We use the JutulDarcy simulator \cite{olav_moyner_sintefmathjutuldarcyjl_2023} for this two-phase flow system. JutulDarcy is implemented in the Jutul framework \cite{olav_moyner_sintefmathjutuljl_2023}, which discretizes spatial fields with finite volumes, discretizes time with an implicit Euler step, and chooses time step sizes automatically with an adaptive time stepper. Jutul solves the discretized system with Newton's method with the necessary Jacobian obtained via automatic differentiation.

\subsection{Observation operator}

The wave dynamics for the seismic operator take place on a time scale up to a few seconds with a resolution on the order of milliseconds.
Changes in the \CO2 saturation occur on a much longer time scale, so we consider the saturation field to be static when simulating seismic measurements.

In line with typical seismic imaging techniques, we construct an approximation of the true pre-injection baseline parameters $m_B$ and $\rho_B$ as a smooth baseline model $m_0$ and $\rho_0$, subtract off the initial seismic measurement, and apply the adjoint Jacobian and a post-processor to obtain a seismic image $A$ as
\begin{align}
    J_0 &= \frac{d\mathcal{H}}{dm_0}(m_0, \rho_0), \label{eq:seismic_baseline_JB} \\
    h(x, \nu \eta) &= PJ_0^T\bigg(\mathcal{H}\big(m(S), \rho(S)\big) + \nu \eta\bigg), \label{eq:seismic_h} \\
    h_B(\nu^*_B \eta^*_B) &= PJ_0^T\bigg(\mathcal{H}(m_B, \rho_B) + \nu^*_B \eta^*_B\bigg), \label{eq:seismic_baseline_h0} \\
    A(x, \nu \eta, \nu_B^* \eta_B^*) &= h(x, \nu \eta) -  h_B(\nu_B^* \eta_B^*), \label{eq:seismic_image}
\end{align}
where the measurements depend only on the saturation of the state tuple $x = (S, P, K)$; $\nu$ and $\nu_B^*$ control the noise magnitude of the simulation and the true observations, respectively; and the post-processor operator $P$ mutes the water layer and scales the image as a function of depth as suggested by \textcite{herrmann_curvelet-based_2009}.
Note that the image $A$ is a constant offset from the observation operator $h$, so it makes no difference in the filter whether we use $h$ or $A$ as the observation.

Each frequency component of the noise is drawn from a zero-centered normal distribution with standard deviation proportional to the corresponding frequency's contribution to the source Ricker wavelet.
We calibrate the noise norm in terms of the signal-to-noise ratio (SNR) $\gamma = \nu^{-2}$, typically expressed in decibels as $10 \log \gamma$.
We consider the ``signal" to be the difference in the observation from the smooth baseline model observation $d_0$.
We scale the noise $\eta$ to have the same norm as the signal so that $\nu \eta$ has the signal-to-noise ratio $\gamma$,
\begin{align}
    \gamma =
    \frac{\|\mathcal{H}(m(S), \rho(S)) - d_0\|^2}{\|\nu\eta\|^2} = \frac{1}{\nu^2}, \label{eq:noise_scaler}
\end{align}
and similarly for $\eta_B^*$, where the norm is defined as $\|\eta\|^2 = \sum_{i,j}\int \eta_{ij}(t)^2 \; dt$, summed over the sources and receivers and integrated over time.

\paragraph*{Linearization}
The Kalman filter is exact for linear transition and observation operators with additive Gaussian noise.
To avoid introducing too much complexity at once to this research, we linearize the seismic operator $\mathcal{H}$.
Thus, we take the initial step of showing that the \enkf/ accommodates the nonlinearity from the two-phase flow transition and the patchy-saturation model.

The seismic operator can be re-parameterized in terms of squared slowness and impedance as $\mathcal{H}_{mz}(m, z)$.
We choose the acoustic impedance $z = \rho / \sqrt{m}$ because the resulting seismic images (computed with inverse scattering imaging conditions) lack low-frequency updates, thereby revealing the shape of the plume better.
We linearize this operator $\mathcal{H}_{mz}$ using the Jacobian $\hat J_0$ with respect to the acoustic impedance about the smooth baseline $z_0 = \rho_0 /\sqrt{m_0}$.
Thus, for the linearized model, we replace $\mathcal{H}(m, \rho)$ in \cref{eq:seismic_h,eq:seismic_image,eq:seismic_baseline_JB,eq:seismic_baseline_h0,eq:noise_scaler} with  
\begin{align}
    \mathcal{\overline H}_{mz}(m, z) = \mathcal{H}_{mz}(m_0, z_0) + \hat J_0 (z - z_0),
\end{align}
where the impedance is computed from the patchy-saturation model as
\begin{align}
z(S; \rho_B, m_B) &= \sqrt{\rho(S) \lambda(S)},
\end{align}
so that the observation operator is 
\begin{align}
    h(x, \nu \eta) &=
    P\hat J_0^T\bigg(\hat J_0 \big(z(S) - z_0\big) + \nu \eta\bigg),
    \label{eq:seismic_linear_h}
\end{align}
with ground-truth observation $y = h(x^*, \nu^* \eta^*)$.

\paragraph*{Uncertain values}
Inverting seismic data itself is difficult, so we do not confound that difficulty with extra uncertainty in the rock physics model.
Instead, we assume $m_B$ and $\rho_B$ are known exactly for the patchy-saturation model, and we compute the smooth baseline model parameters from a Gaussian blur of $m_B$ and $\rho_B$.

\paragraph*{Discretization and solution}
We use the JUDI software \cite{witte_large-scale_2019,mathias_louboutin_slimgroupjudijl_2024} for this seismic system, which solves the wave equation using the Devito package \cite{louboutin_devito_2019,luporini_architecture_2020}.
The method is an 8th order spatial finite differencing scheme with 2nd order in time with time step size chosen based on the CFL conditions described by \textcite{lines_recipe_1999}.

\subsection{Comparison to other algorithms}

For the \enkf/, we initialize $\es$ ensemble members with different permeability models and zero saturation.
These states represent possibilities for the state of the reservoir at the initial time.
For each ensemble member, we simulate two-phase flow with \CO2 injection until the measurement time.
The \enkf/ uses the ensemble covariance and the measurement covariance to update the saturations of each ensemble member towards a value consistent with the observations, according to \cref{eq:enkf_update}.
Then the simulate-update process is repeated, with the two-phase flow simulation advancing each ensemble member until the next observation time followed by a measurement being incorporated that updates the ensemble states.

The \enkf/ assimilates data from simulations and observations. In order to be worth the extra computational effort, the \enkf/ must perform better than similar algorithms that use only data from simulations or only data from observations.
We propose a comparison with two baseline ensemble methods of updating the forecast, referred to as no-observations method \emph{NoObs} and a just-observations method \emph{JustObs}.
The NoObs method does not incorporate observations; it represents the best prediction we can obtain based on prior physics knowledge, specifically, the two-phase flow model and the probability distribution describing the permeability field.
The JustObs method does not use the fluid-flow physics knowledge; it represents the best prediction we can obtain based solely on observations, specifically, waveform observations from \cref{eq:fullwaveform_H}.
The JustObs optimization is ill-posed and can be sensitive to the initial guess for optimization, so we initialize it with the forecasted ensemble mean and add a regularization term $C(x)$, defined later.
The uninitialized and unregularized version performed much worse, and we do not show it.
In that sense, JustObs does use the transition physics knowledge, but much less than the \enkf/ and NoObs methods.

The forecast for each method is computed in the same way. The previous analysis state $x_a$ at time step $n-1$ for ensemble member $i$ is simply advanced forward in time with the transition operator, written as
\begin{align}
x_{f,i}^{n} &= f(x_{a,i}^{(n-1)}).
\end{align}

We can see the similarities between three methods by writing them each as an optimization, dropping the subscript from $x$ since they are all at time step $n$:
\smallskip

\noindent For NoObs,
\begin{align}
\begin{split}
\quad  x_{a,i} &= \arg \max_{x} p(x|x_{f,i}^{0})
\\&= \arg \min_x \|x - x_{f,i}\|_{B_0^{-1}}^2.
\end{split}\\
\intertext{For JustObs,}
\begin{split}
\quad  x_{a,i} &= \arg \max_{x} p(x | y^{n})
\\&= \arg \min_x \|h(x) - y^{n}\|_{R^{-1}}^2 + C(x).
\end{split}\\
\intertext{For \enkf/,}
\begin{split}
\quad  x_{a,i} &= \arg \max_{x} p(x | y^{1:n})
\\&= \arg \min_x \|\hat h(x) - y^{n}\|_{R^{-1}}^2 + \|x - x_{f,i}\|_{B_0^{-1}}^2.
\end{split}
\end{align}
Recall from \cref{eq:enkfopt} that the \enkf/ uses a linearized observation operator (written $\hat h(x)$ here) based on the ensemble members and constrains $x$ to be expressed by the range of the ensemble mean deviations. The version here is simplified to more easily compare to NoObs and JustObs.

Because these are ensemble methods, each ensemble member has a separate realization for $x_{f,i}$ and $x_{a,i}$. Each of the above equations is solved for each ensemble member.
Since each ensemble member has a different $x_{f,i}$, the NoObs and \enkf/ methods include uncertainty in the resulting $x_{a,i}$.
The NoObs optimization has a simple solution $x_{a,i} = x_{f,i}$ because the observations are not included.
In JustObs, the resulting $x_{a,i}$ is identical for each ensemble member because $x_{f,i}$ does not appear in the optimization expression.
For all methods, due to each ensemble member's unique permeability field, the ensemble states tend to drift away from each other and from the ground-truth state when the transition function is applied.

For the JustObs optimization, we use a projected quasi-Newton algorithm with spectral projected gradient algorithm described by \textcite{schmidt_optimizing_2009} and implemented in Julia by \textcite{mathias_louboutin_slimgroupslimoptimjl_2024}.
For regularization $C(x)$, we use $C(x) = \| \lambda_h^{-1} L_h S\| + \| \lambda_v^{-1}L_y S\|$, with length scales $\lambda_h$ and $\lambda_v$ and linear operators $L_h$ and $L_v$ that compute the horizontal and vertical gradients of the saturation field $S$ with finite differences. We experimented with the following norms to regularize the gradient of the saturation: $\ell_1$, known as total variation (TV) regularization \cite{rudin_nonlinear_1992}; $\ell_2$, known as Tikhonov regularization \cite{tikhonov_solutions_1977}; and a hybrid $\ell_1$/$\ell_2$ norm \cite{bube_hybrid_1997} that benefits from the sparsity behavior of the $\ell_1$ norm without requiring the $\ell_1$-norm projection.
These norms have successfully been used for inverting seismic velocity.
A useful albeit simplified characterization of these regularizations is TV yields piecewise-constant solutions, Tikhonov yields smoothly-varying solutions, and the hybrid yields piecewise-smooths solutions.
The \CO2 plume should be piecewise-smooth instead of piecewise-constant, so we expect the hybrid regularization to yield better results.
Indeed, we empirically find the hybrid performs slightly better, but the optimization with each of these norms gave very poor estimates of \CO2 saturation, with nonzero \CO2 saturation placed across much of the domain even with non-noisy data and perfect velocity models.
More work could be done to improve the JustObs algorithm, e.g., based on \cite{modrak_seismic_2016}. However, since the focus of this paper is the \enkf/, comparing a simple JustObs algorithm to the simplest \enkf/ algorithm is reasonable.

\subsection{Noise tests}

We additionally test a range of values for the parameters for the \enkf/ while keeping the transition and observation models fixed.
We show the \enkf/'s resiliency to deviations from the ideal parameters, and we provide guidance on how to choose some parameters.
Specifically, the \enkf/ algorithm has choices for how the noise is handled and estimated, and we are interested in how the accuracy of the \enkf/'s state estimate accuracy changes under incorrect assumptions and different ways to handle the noise.

Recall that the Kalman update can be written in terms of covariances, and the \enkf/ uses sample covariances from the ensemble, denoted here with $\widehat{\cov}$ with observation noise covariance estimate or regularization $R$,
\begin{align}
x_{a,i} &= x_{f,i} + \widehat{\cov}(x_f, y_f)(\widehat{\cov}(y_f) + R)^{-1}(y - y_{f,i}).\label{eq:noise_tests_enkf2}
\end{align}
Our seismic data is represented as $y_{f,i} = h(x_{f,i}, \nu \eta_i)$, where the norm of $\eta_i$ is fixed and $\nu$ determines the SNR (expressed as $-20 \log \nu$ dB).
If the noise covariance is known precisely, $\widehat{\cov}(y_f)$ should be computed without noise as $\widehat{\cov}(h(x_f, 0))$, and $R$ should fully represent the noise covariance.

However, we don't generally know the covariance of the noise, so if we instead compute $\widehat{\cov}(y_f)$ with the noise as $\widehat{\cov}(h(x_f, \nu \eta_h))$, then $\widehat{\cov}(y_f)$ implicitly contains an estimate of the noise covariance. Unfortunately, the number of ensemble members is much smaller than the observation size, which makes $\widehat{\cov}(y_f)$ singular. This necessitates regularization in the inversion of the sample observation covariance, in which case, the matrix $R$ is ideally close to the true noise covariance only in the directions that are not already sampled in $\widehat{\cov}(y_f)$.

Typical handling of the regularization simply parametrizes $R$ as a diagonal scaling of the identity, and we follow suit here. The noise variance is proportional to $\nu^2$, so we parametrize $R$ as $R(\nu, \beta) = \nu^2 \beta^2 I$, with scaling parameter $\beta^2$ being the estimate of the average variance of the observation vector's entries when the SNR is 0 dB. If we have the true observation noise covariance $R^*$, we would choose $\beta^2 = \text{mean}(\text{diag}(R))$.

For our noise tests, we consider changing the four noise parameters: first, $\beta$, which determines the amount of regularization in the inversion of $\widehat{\cov}(y_f)$; then $\alpha \in \{0, 1\}$, a binary value which determines whether to simulate the noise in the estimate of $\widehat{\cov}(y_f)$; and finally, $\nu$ and $\nu^*$, which determine the magnitude of the noise in the simulation and the synthetic ground-truth observations. These scalar parameters are shown underlined in the update equation as
\begin{multline}
\labelAndRemember{eq:noisy_tests_enkf_update}{
    x_{a,i} = x_{f,i} +
    \widehat{\cov}\big(x_f, h(x_f, \underline{\nu} \eta)\big)
\bigg(
    \widehat{\cov}(h(x_f, \underline{\alpha \nu} \eta))
    + \underline{\nu^2 \beta}^2 I
\bigg)^{-1}
\big[
    h(x^*, \underline{\nu^*} \eta^*) -
    h(x_{f,i}, \underline{\nu} \eta_i)
\big]}.
\end{multline}
We refer to changing $\beta$ and $\alpha$ as \emph{regularization tests}, changing $\nu$ as \emph{simulated noise tests}, and changing $\nu^*$ as \emph{true noise tests}, described in more detail below.

\paragraph*{Regularization tests} The regularization tests measure performance with different values of $\beta$ and $\alpha$, keeping $\nu = \nu^*$ fixed. As $\beta$ is increased, the update to $x_k$ goes to 0, so we expect the \enkf/ estimates to approach the NoObs estimates.
For $\alpha = 1$, the noise is included in the samples for the sample covariance, and $\beta$ is simply a regularization parameter.
For $\alpha = 0$, $R$ should represent the covariance of the noise in the observation operator, and $\beta$ is chosen to make the resulting variance match the noise covariance on the diagonal.

\paragraph*{Simulated noise tests} For these tests, we measure performance when the noise estimate $\nu \ne \nu^*$ while keeping $\beta$, $\alpha$, and $\nu^*$ fixed.
If $\nu$ equals the true noise $\nu^*$, then our ensemble's observation noise variance is an unbiased estimate of the true observation noise variance.
When the noise estimate is too high, we expect to regress to the NoObs results, and when the noise estimate is too low, we expect to fit the noisy data too closely.
In practice, the true noise variance is not known exactly but can be estimated based on observations and technical specifications of the sensors.
These tests show the robustness of the \enkf/ to mis-calibration of the noise.
Note that we consider strictly unbiased (zero-mean) noise.
Biased errors typically occur from incorrect forward models, but we do not address those here.

\paragraph*{True noise tests} Finally, we further show the robustness of the \enkf/ by measuring performance with different values of the true noise magnitude $\nu^*$.
We are especially interested in how the performance decays for larger values of noise.
For these tests, we keep $\nu = \nu^*$ and keep $\beta$ and $\alpha$ fixed.

\makeatletter
\newlength \singlecolumnwidth
\if@twocolumn
  \setlength \singlecolumnwidth {0.9\columnwidth}
\else
  \setlength \singlecolumnwidth {0.49\linewidth}
\fi
\makeatother

\begin{figure}[htbp]
\centering
\begin{minipage}[c]{\singlecolumnwidth}
    \includegraphics[width=\linewidth]{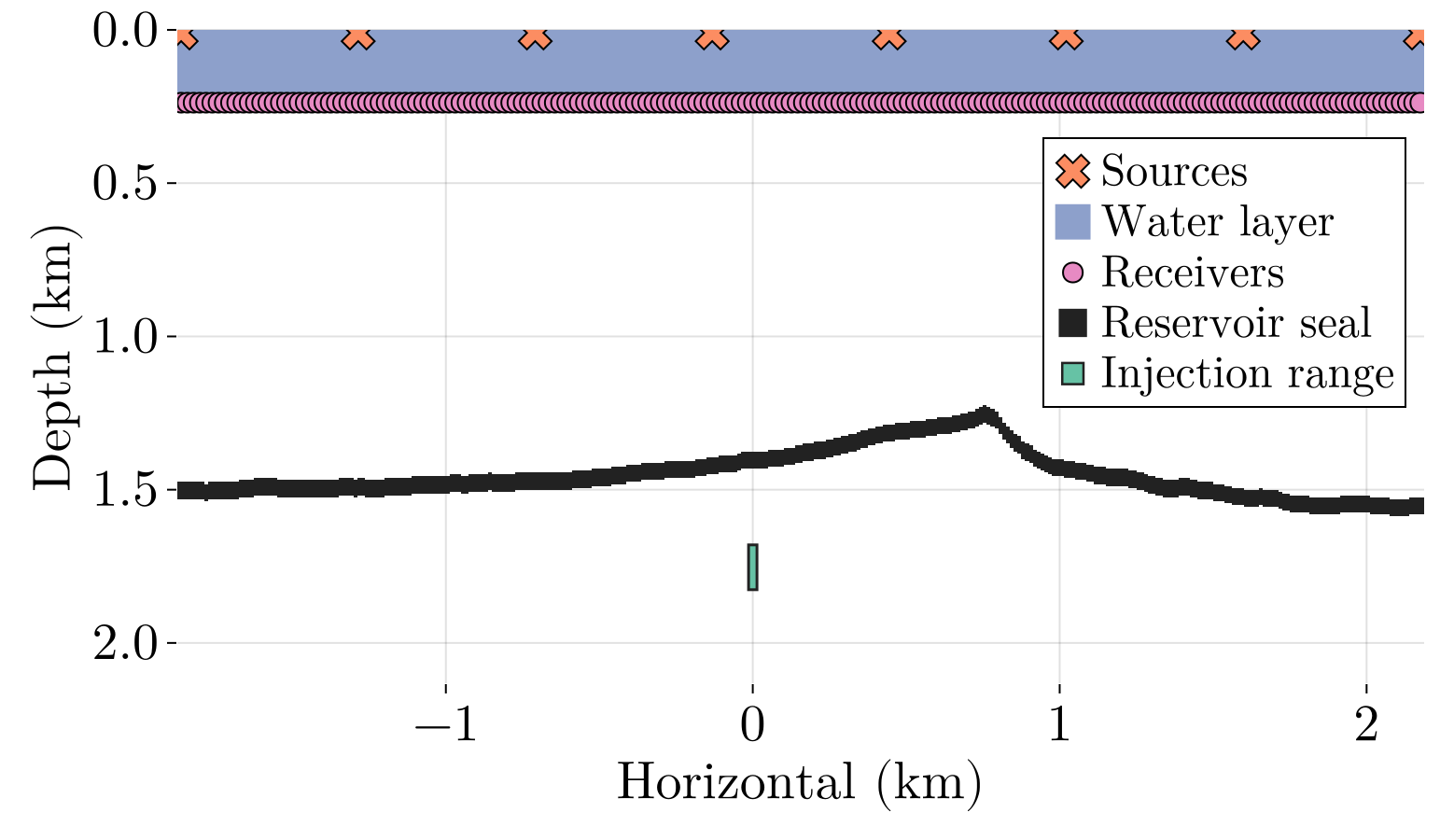}
\end{minipage}
\caption{Simulation domain}
\label{fig:experiment-setup}
\end{figure}

\begin{table}[htbp]
\centering
\begin{subtable}{\singlecolumnwidth}
    \centering
    \begin{tabular}{r|l}
        Domain size & 4.05 km $\times$ 2.125 km \\
        Grid size & 325 $\times$ 341 \\
        Cell size & 12.5 m $\times$ 6.25 m
    \end{tabular}
    \caption{Shared domain parameters.}
    \label{tab:domain-params}
\end{subtable}
\bigskip\\
\begin{subtable}{\singlecolumnwidth}
    \centering
    \begin{tabular}{r|l}
        Simulation length & 5 years \\
        Injection depth & 1.76 km \\
        Injection extent & 37.5 m \\
        Injection rate & 0.8 Tg/year \\
        Residual saturation & 0.1 \\
        $K_v / K_h$ & 0.36
    \end{tabular}
    \caption{Plume parameters.}
    \label{tab:plume-params}
\end{subtable}
\bigskip\\
\begin{subtable}{\singlecolumnwidth}
    \centering
    \begin{tabular}{r|l}
        Simulation length & 1.8 s \\
        Source dominant frequency & 24 Hz \\
        Source maximum amplitude & 7.8 MPa \\
        Signal-to-noise ratio & $8 \text{ dB} $\\
        Number of receivers & 200 \\
        Number of sources & 8
    \end{tabular}
    \caption{Seismic parameters.}
    \label{tab:seismic-params}
\end{subtable}
\caption{Simulation parameters}
\label{tab:params}
\end{table}

\section{Test scenario}

We show the ensemble Kalman filter performance on a synthetic \CO2 injection problem based on the Compass model, which is a synthetic benchmark for seismic full waveform inversion \cite{e_jones_building_2012}.
We simulate the \CO2 injection for 5 years, with new seismic observations every year.
The simulation domain is shown in \cref{fig:experiment-setup}, and the scalar simulation parameters are shown in \cref{tab:params}.

\begin{figure}[htbp]
\centering
\begin{subfigure}{\singlecolumnwidth}
    \centering
    \includegraphics[width=\linewidth]{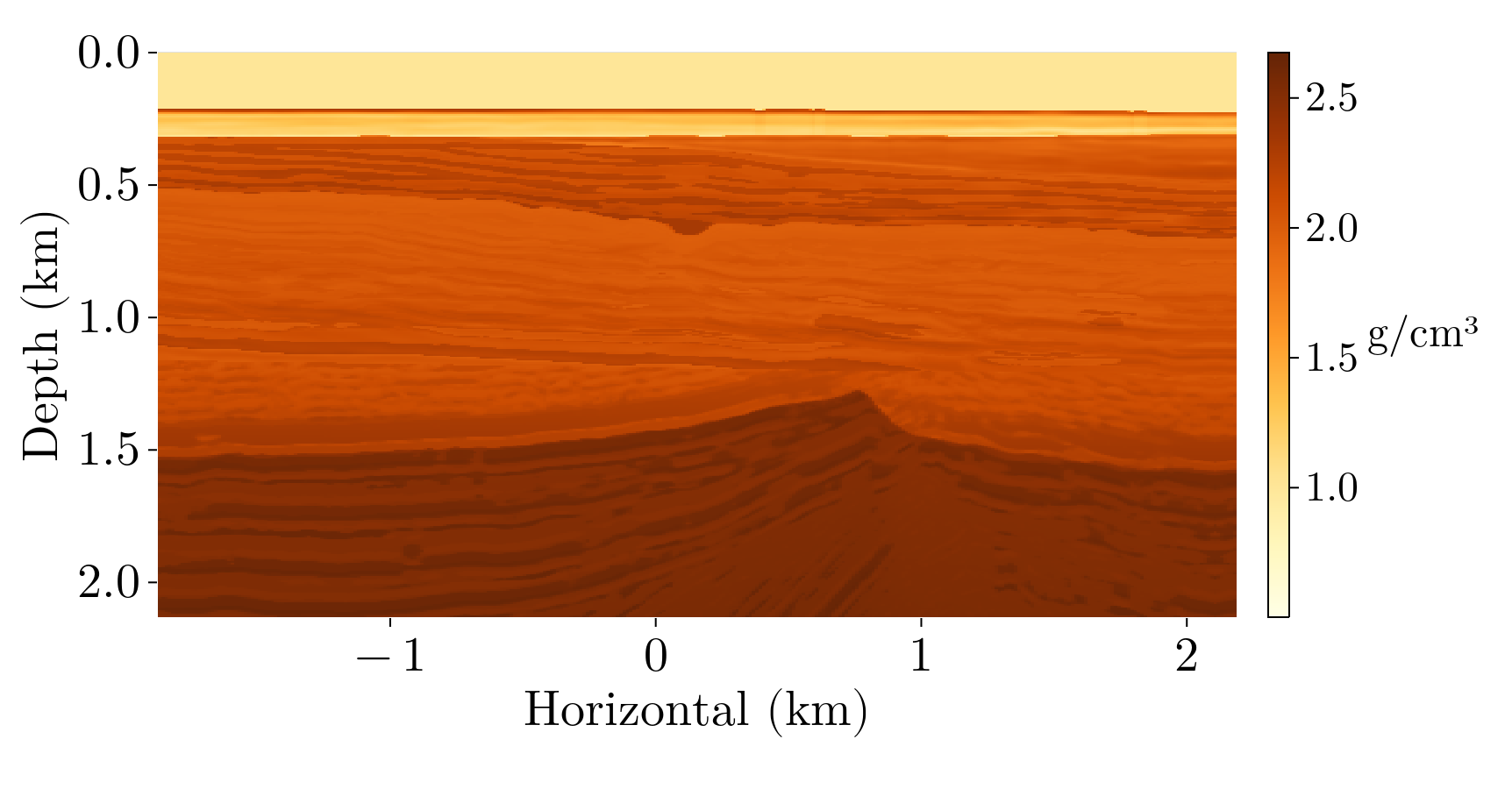}
    \caption{Density field}
\end{subfigure}
\begin{subfigure}{\singlecolumnwidth}
    \centering
    \includegraphics[width=0.93\linewidth]{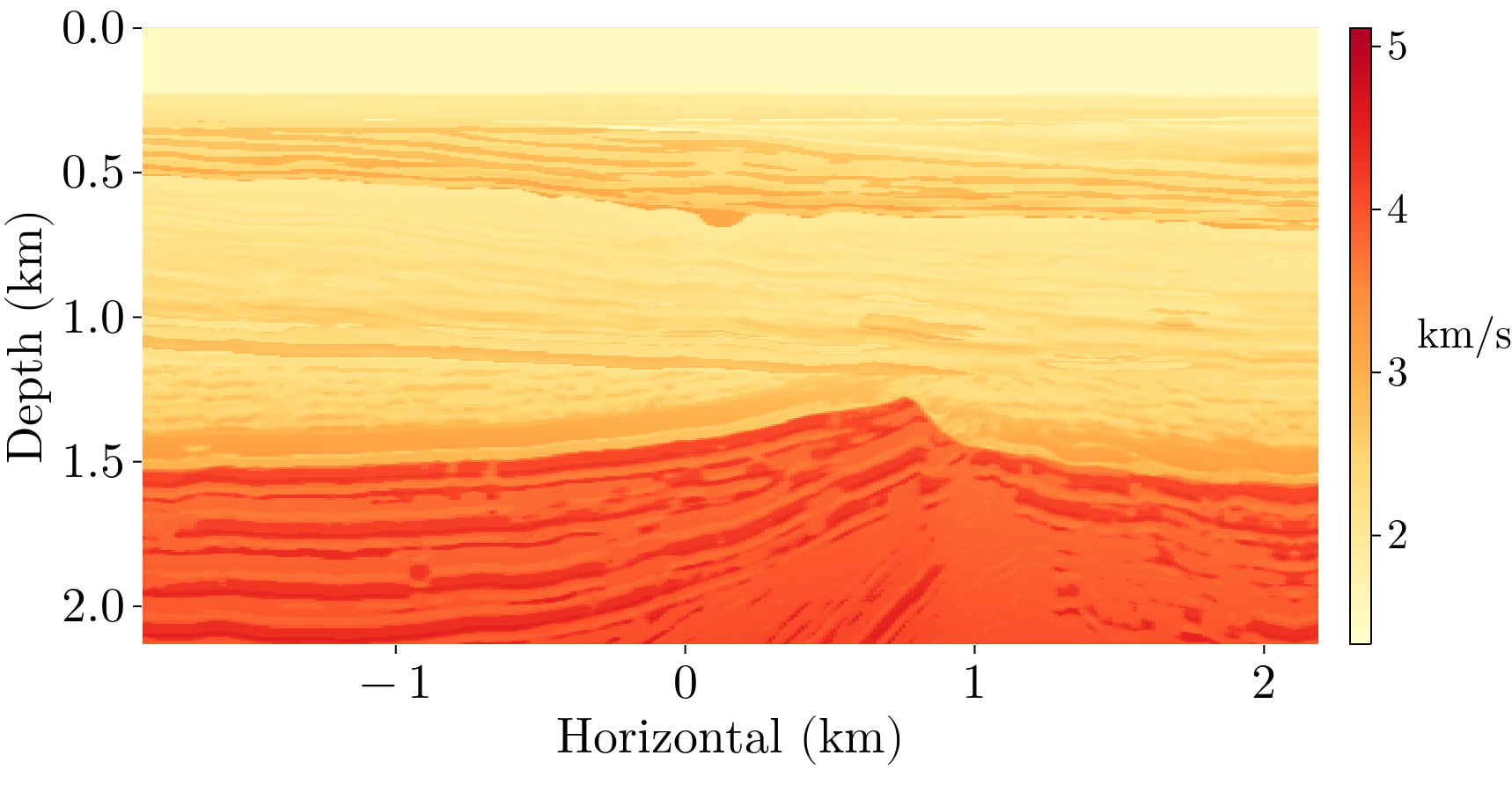}
    \caption{P-wave velocity field}
\end{subfigure}
\caption{Seismic field parameters. These come from a 2D slice of the compass model.}
\label{fig:seismic-parameters}
\end{figure}

The Compass model is a large 3D model designed to capture the geological complexities of seismic behaviors in real systems. Running simulations with the full model is expensive, so we limit the domain to a 2D slice, shown in \cref{fig:seismic-parameters}.

For each ensemble member, the permeability is a function of a deformed version of the true velocity field $v$. The applied distortion is an elastic deformation provided by the Augmentor.jl software \cite{bloice_augmentor_2017}.
Let $L_i$ be the domain extent in either the vertical or horizontal direction.
The deformation is expressed as a regular coarse grid of displacement vectors sampled uniformly with components between $-L_i$ and $L_i$ and then normalized such that the 2-norm over the grid along each displacement component of the vectors is $0.2L_i$. For a displacement field $\vec{d}(\vec{r})$, the deformed velocity field is expressed as $v'(\vec{r}) = v(\vec{r} + \vec{d}(\vec{x}))$.

The dimensions of the coarse grid are chosen uniformly between 30 and 50, and the distortion vectors along the boundary are set to 0.
Furthermore, a Gaussian kernel with length scale between $L_i/15$ and $L_i/25$ is applied to smooth the distortion field.
For our grid, the root-mean-squared distortion is 20.3 meters in the horizontal direction and 10.7 meters in the vertical direction. After deforming the velocity field, the resulting permeability field using \cref{eq:VtoK}.

Each ensemble member has a different permeability model.
To avoid extremely high pressure, we choose the simulated injection depth for each permeability model to be the maximum permeability value in the 120 meter range labeled in \cref{fig:experiment-setup}, and we initialize the saturation field with a random-valued saturation patch of 57 grid cells around that location.

\begin{figure*}[htbp]
\centering
\includegraphics[width=\textwidth]{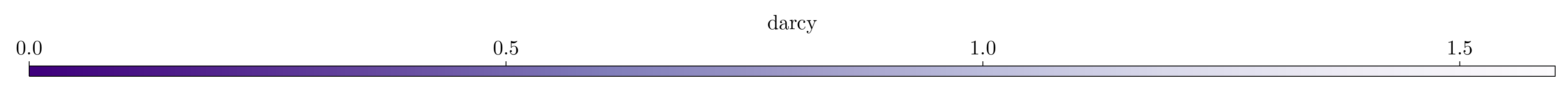}
\bigskip
\begin{subfigure}{\textwidth}
    \centering
    \includegraphics[width=\textwidth]{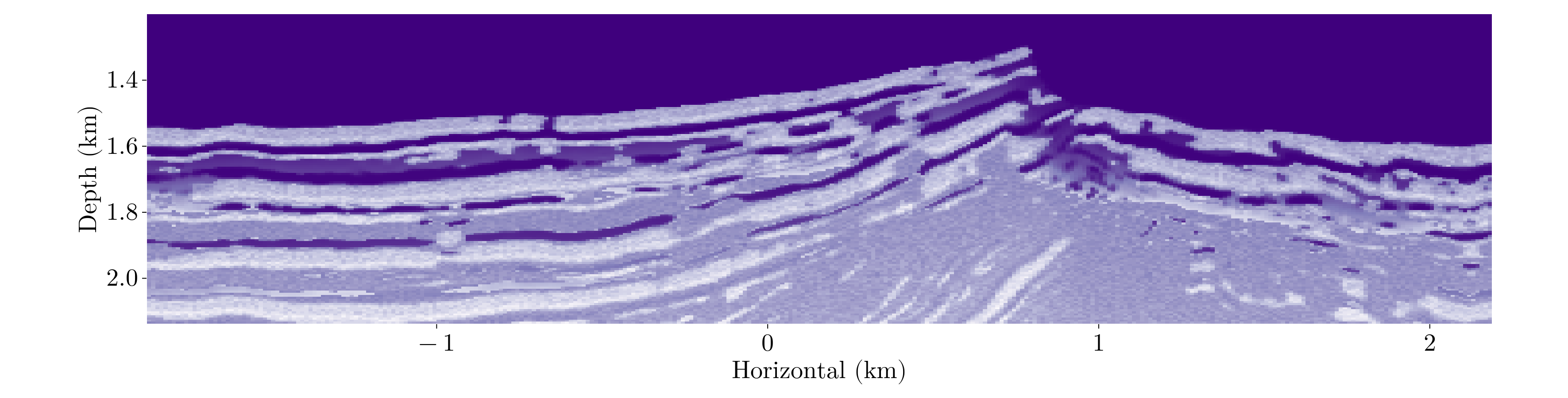}
    \caption{Ground-truth permeability field}
    \label{fig:permeability-ground-truth}
\end{subfigure}
\bigskip
\begin{subfigure}{\textwidth}
    \centering
    \includegraphics[width=\textwidth]{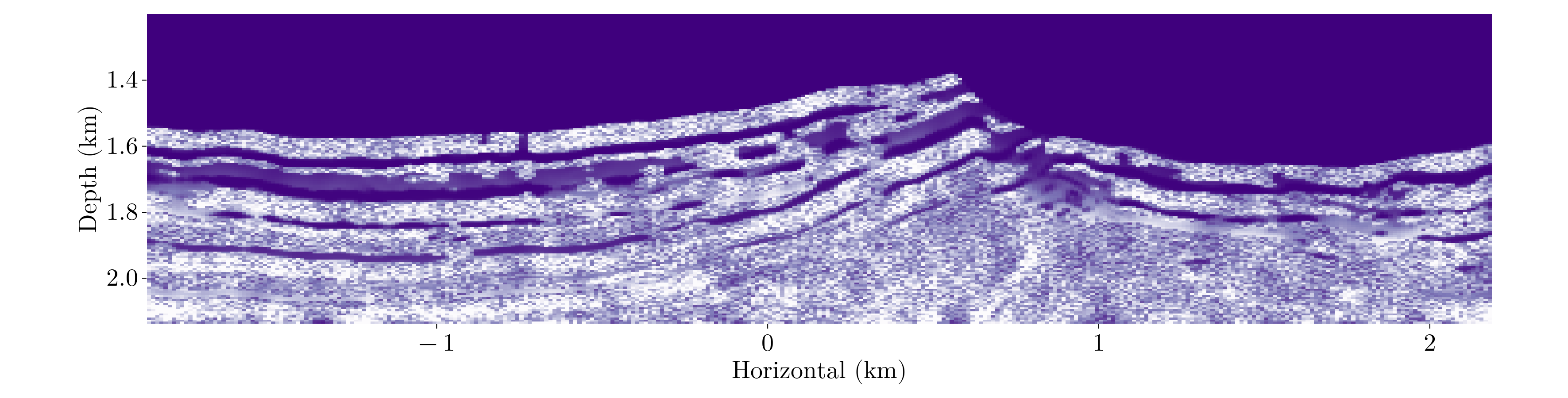}
    \caption{Example permeability sample for ensemble}
\end{subfigure}
\bigskip
\begin{subfigure}{\textwidth}
    \centering
    \includegraphics[width=\textwidth]{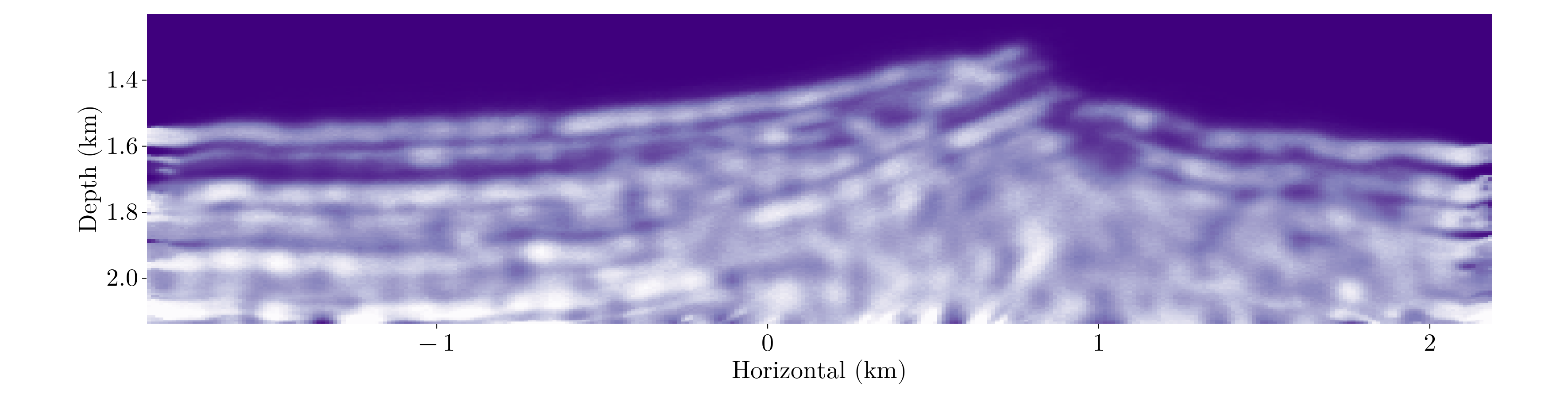}
    \caption{Mean permeability field for ensemble}
\end{subfigure}
\begin{subfigure}{\textwidth}
    \centering
    \includegraphics[width=\textwidth]{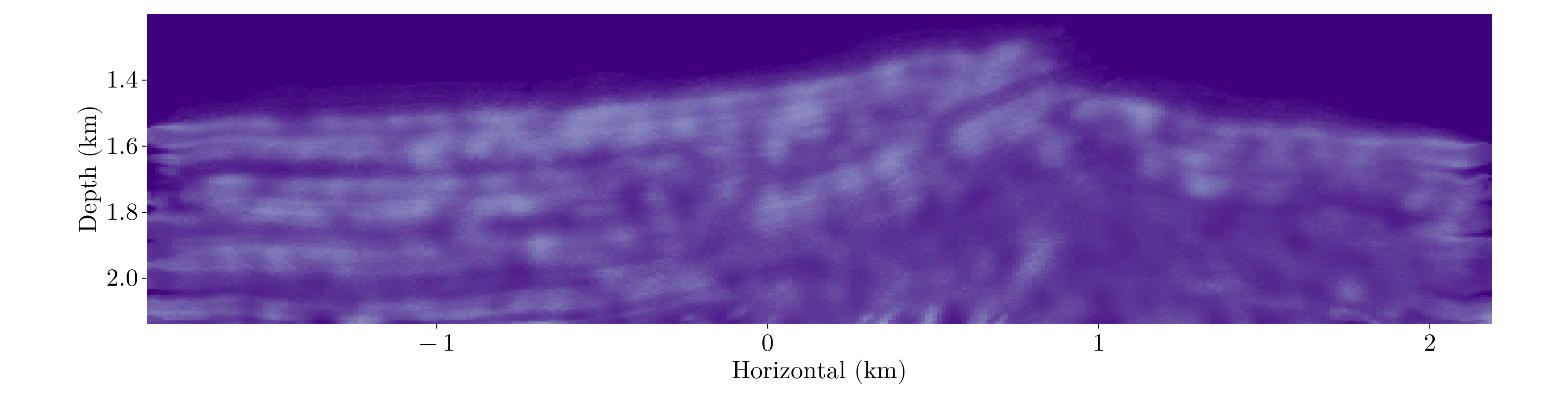}
    \caption{Standard deviation of permeability fields}
\end{subfigure}
\caption{The permeability field is computed from the velocity field with additional perturbations. The ground truth more closely resembles the velocity field, while ensemble permeabilities are much more noisy.}
\label{fig:permeability}
\end{figure*}

\begin{figure*}[htbp]
\centering
\includegraphics[width=\textwidth]{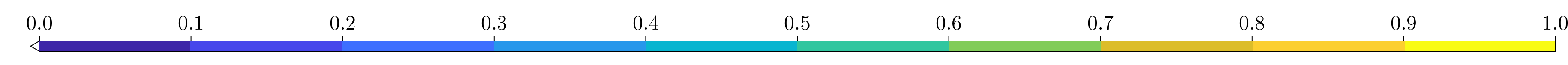}
\begin{subfigure}{\textwidth}
    \centering
    \includegraphics[width=\textwidth]{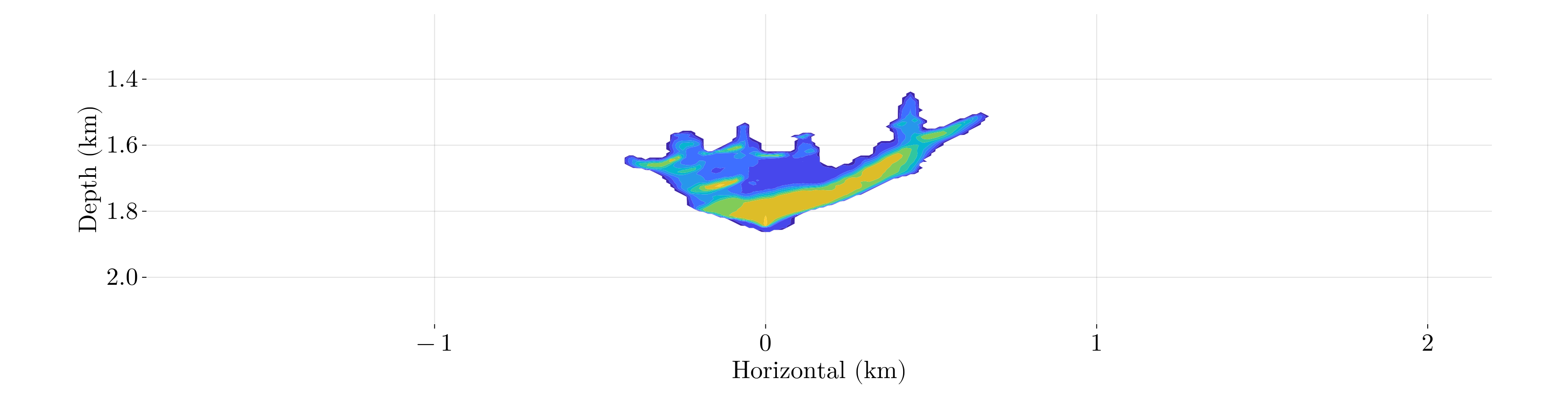}
    \caption{\CO2 saturation after 1 year. Zeros are transparent.}
\end{subfigure}
\bigskip
\begin{subfigure}{\textwidth}
    \centering
    \includegraphics[width=\textwidth]{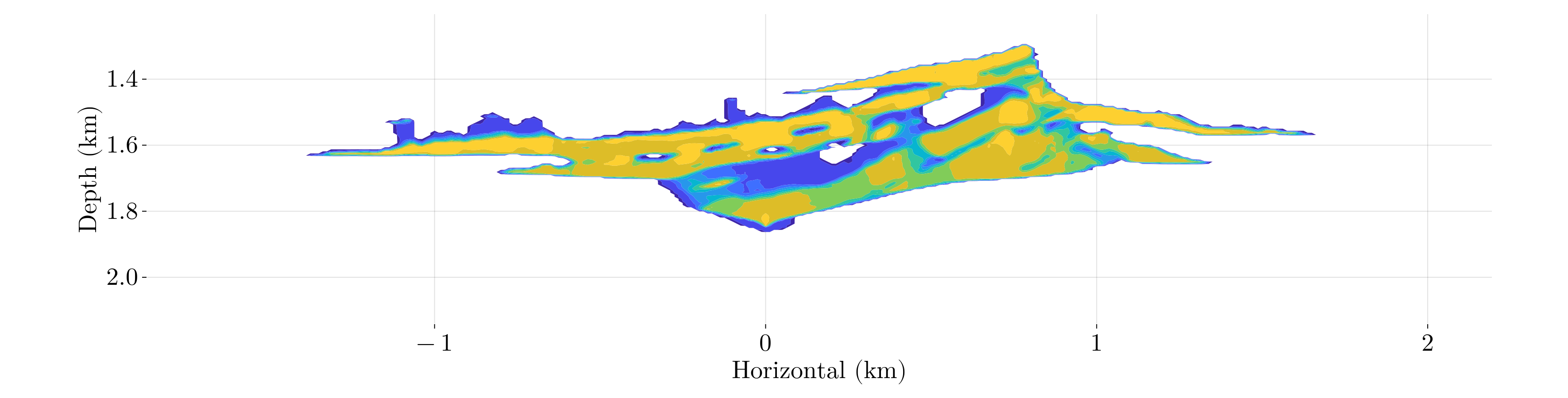}
    \caption{\CO2 saturation after 5 years.}
\end{subfigure}
\bigskip
\settoheight{\imageheight}{\includegraphics[width=0.43\linewidth]{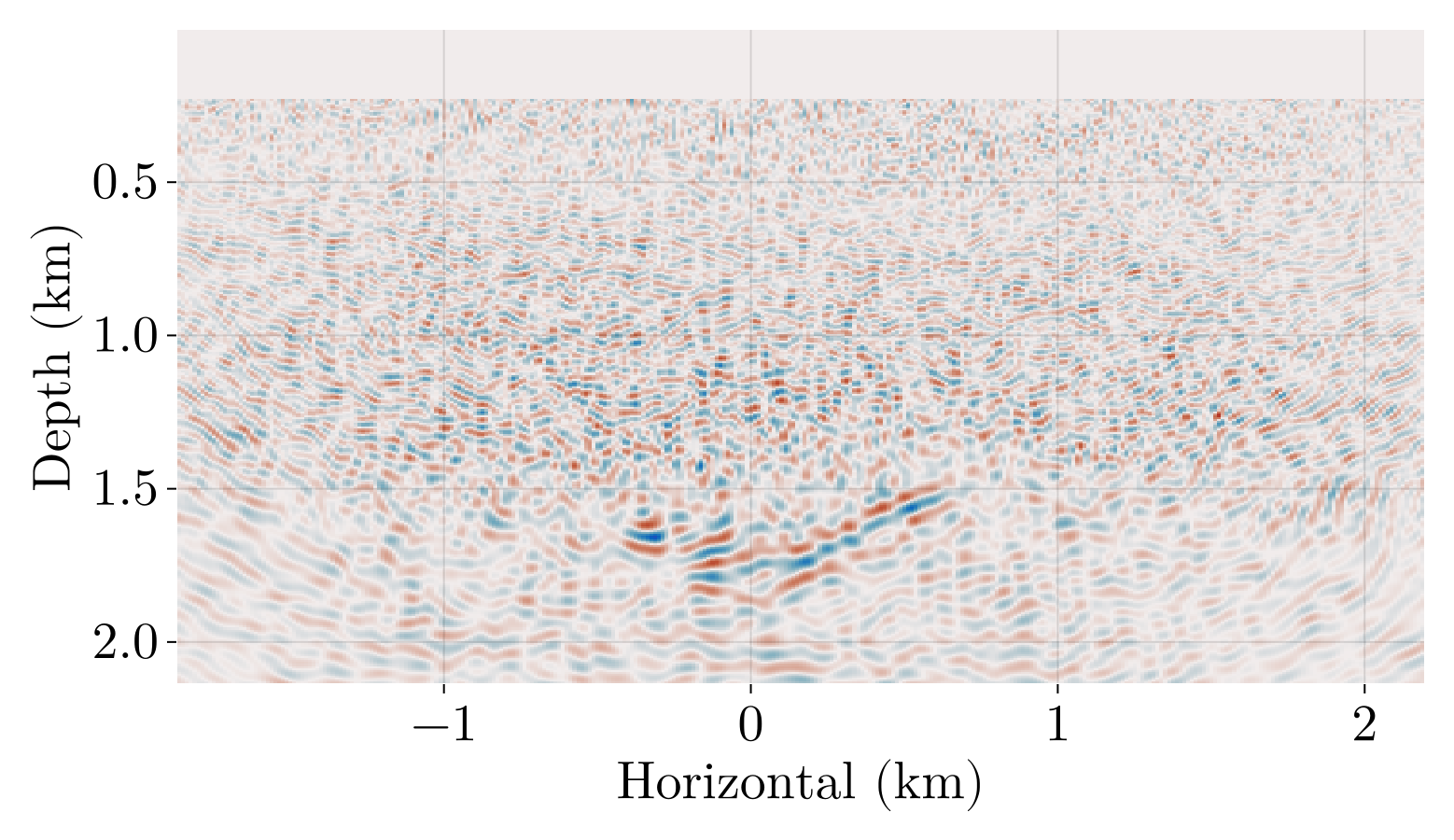}}
\begin{subfigure}{0.43\textwidth}
    \centering
    \includegraphics[height=\imageheight]{figs/ground_truth/rtm/rtms_offset_noisy_cb_balance-0006.png}
    \caption{Observation (1 year)}
\end{subfigure}
\begin{subfigure}{0.53\textwidth}
    \centering
    \includegraphics[height=\imageheight]{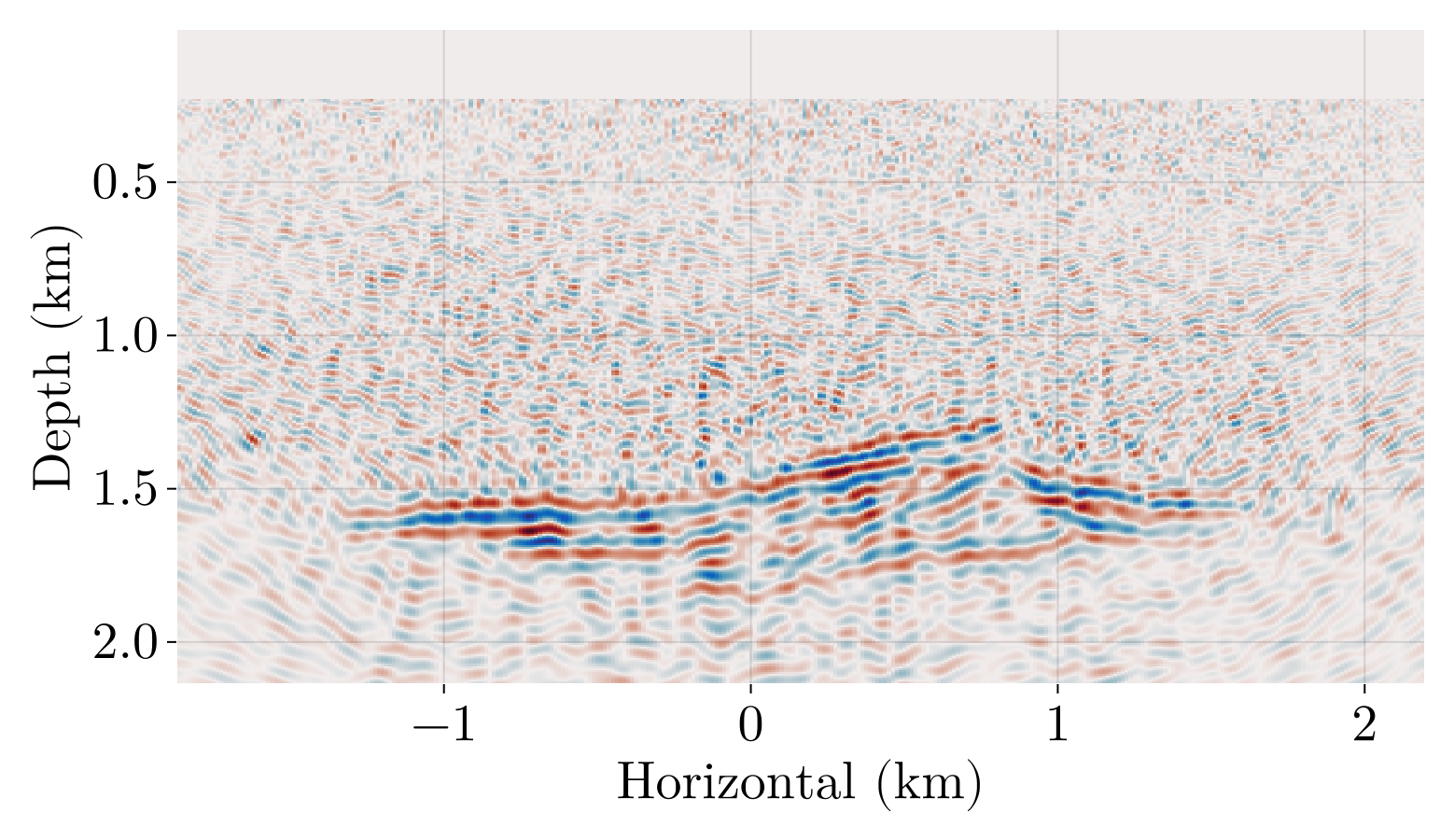}%
    \includegraphics[height=\imageheight]{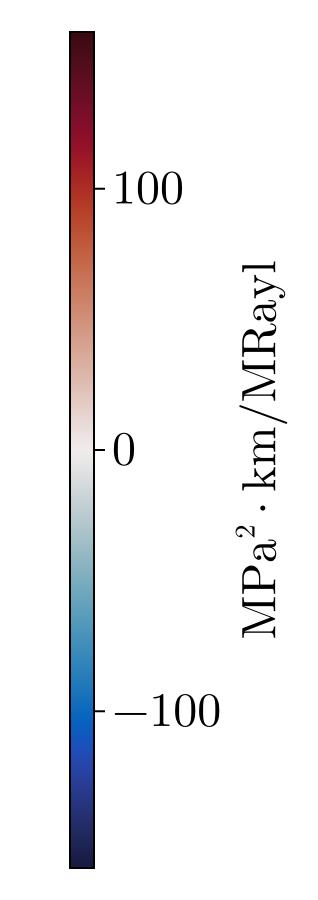}
    \caption{Observation (5 years)}
\end{subfigure}
\caption{
\CO2 plume evolution over time. Observation units for RTM are squared pressure per impedance times depth,  with pressure in megapascals, impedance in megarayls ($\text{g/cm}^{3}\cdot\text{km/s}$), and depth in kilometers.
}
\label{fig:ground-truth-plume}
\end{figure*}

\Cref{fig:permeability} shows the ground truth permeability, an example ensemble member's permeability, as well as the mean and standard deviation of the 256 permeability samples used for the ensemble. \Cref{fig:ground-truth-plume} shows the ground-truth saturation and observation at two times.

\section{Numerical results}

First, we define the metrics we use to measure performance.
Since this is a synthetic experiment, we benefit by being able to compare each saturation estimate $S$ directly to the ground truth saturation $S^*$.
The ensemble mean is the standard estimate to show and is the best estimate of the true saturation field in terms of the $\ell_2$ norm assuming a Gaussian distribution.
Since our transition and observation operators are nonlinear, our distribution is not Gaussian.
Still, there is not a clear choice for a different function of the ensemble states to use, so we show the mean and compute error statistics using the mean.
We use root-mean squared error (RMSE) defined as
\begin{align}
    \RMSE(S, S^*) &= \sqrt{\frac{1}{|\Omega|} \int_\Omega (S(\vec{r}) - S^*(\vec{r}))^2 \; d\vec{r}},
\end{align}
approximated as the mean of the squared error in each grid cell of the simulation domain $\Omega$.
We also computed the structural similarity index measure (SSIM) metric defined in \cite{wang_image_2004}.
SSIM takes into account edges as well as values and is strongly correlated with how humans perceive similarity between two images.
For our tests, we found the SSIM metric to not yield any more information than the RMSE, so we show only RMSE for most plots.

\subsection{\enkf/ compared to baselines}

\begin{figure*}[htbp]
\centering
\includegraphics[width=\textwidth]{figs/ground_truth/plume/saturation_colorbar.png}
\bigskip
\begin{subfigure}{\textwidth}
    \centering
    \includegraphics[width=\textwidth]{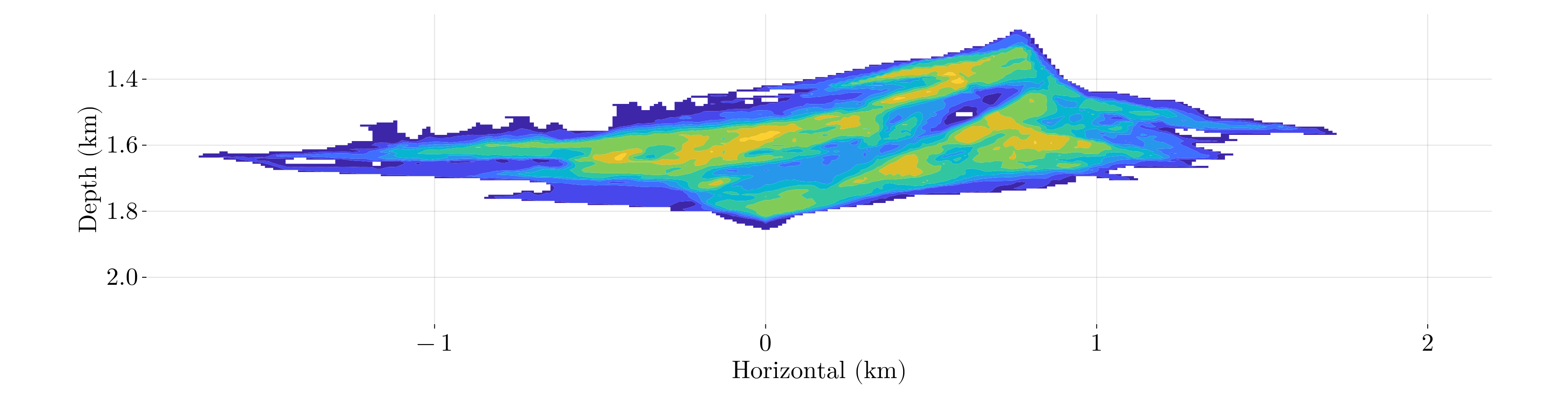}
    \caption{NoObs prediction with 0 measurements}
\end{subfigure}
\bigskip
\begin{subfigure}{\textwidth}
    \centering
    \includegraphics[width=\textwidth]{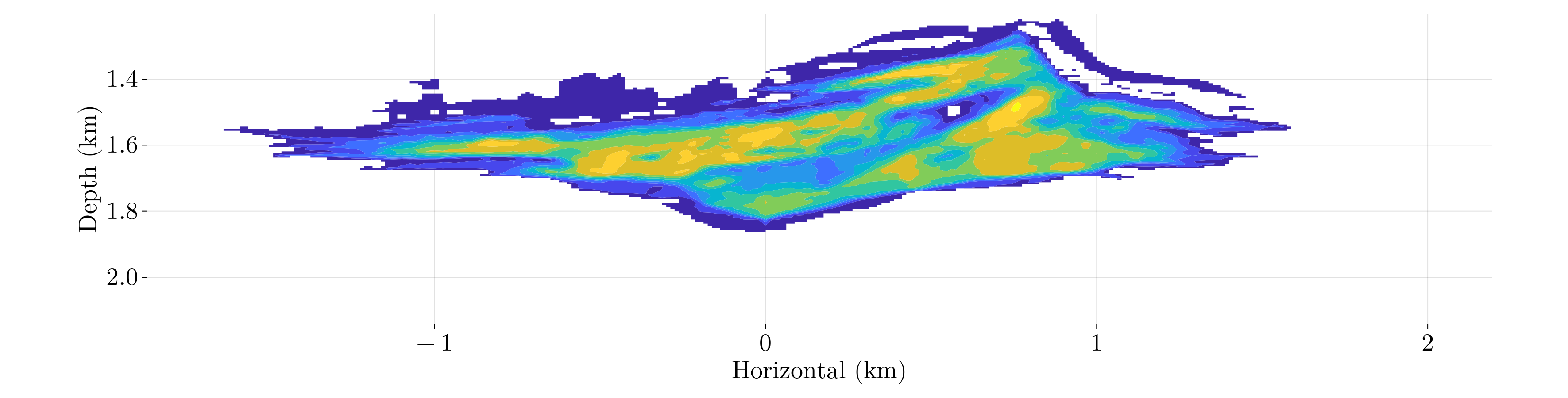}
    \caption{\enkf/ prediction with 4 measurements}
\end{subfigure}
\bigskip
\begin{subfigure}{\textwidth}
    \centering
    \includegraphics[width=\textwidth]{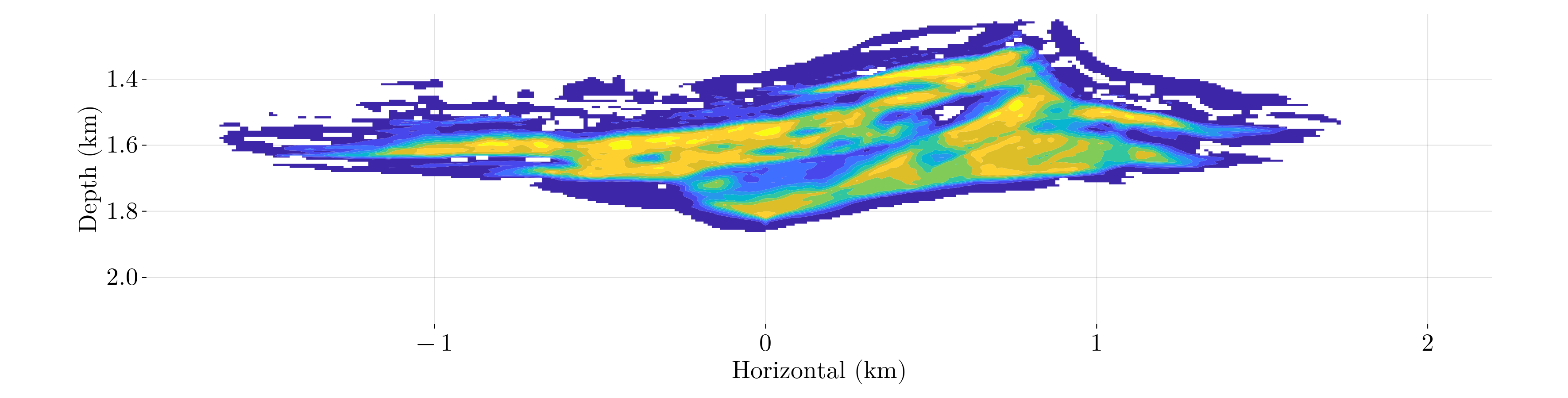}
    \caption{\enkf/ analysis with 5 measurements}
\end{subfigure}
\bigskip
\begin{subfigure}{\textwidth}
    \centering
    \includegraphics[width=\textwidth]{figs/ground_truth/plume/saturation-0026.png}
    \caption{Ground truth}
\end{subfigure}
\caption{\CO2 saturation after 5 years for NoObs, \enkf/, and ground-truth.}
\label{fig:alg-plumes}
\end{figure*}

\begin{figure*}[htbp]
\centering
\includegraphics[width=\textwidth]{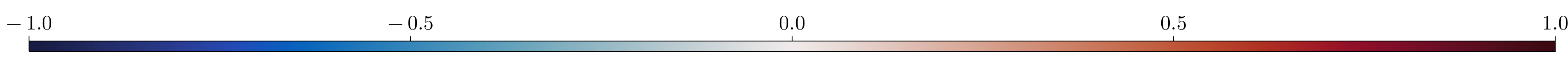}
\bigskip
\begin{subfigure}{\textwidth}
    \centering
    \includegraphics[width=\textwidth]{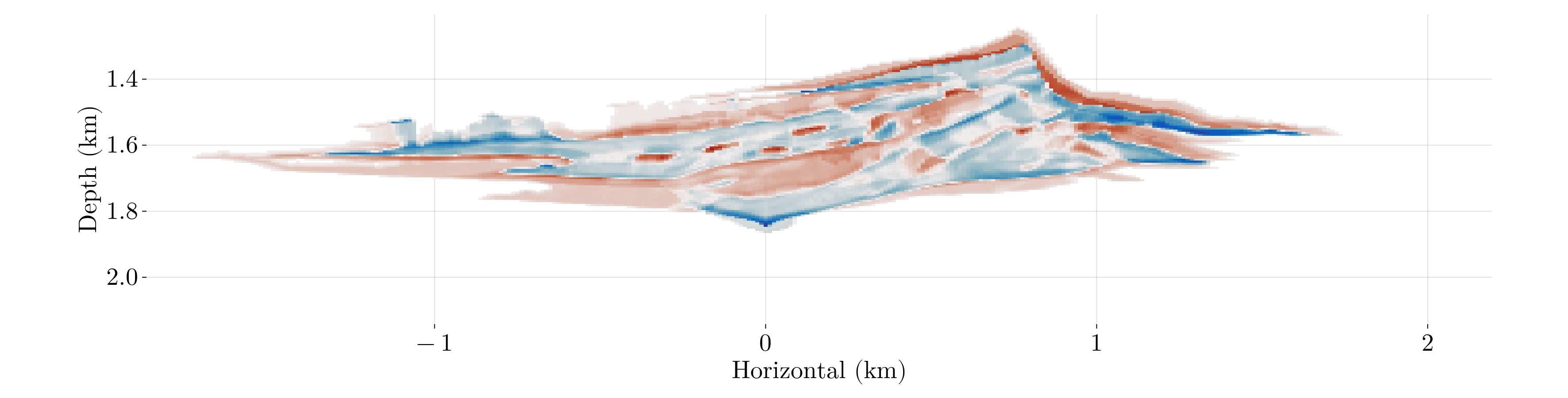}
    \caption{NoObs prediction error}
\end{subfigure}
\bigskip

\begin{subfigure}{\textwidth}
    \centering
    \includegraphics[width=\textwidth]{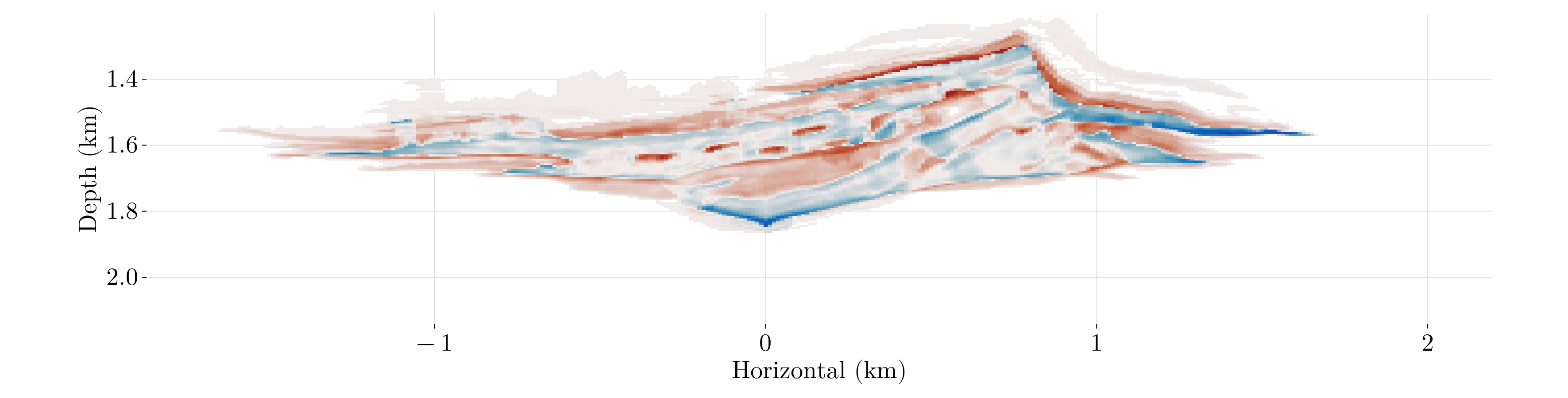}
    \caption{\enkf/ prediction error}
\end{subfigure}
\bigskip

\begin{subfigure}{\textwidth}
    \centering
    \includegraphics[width=\textwidth]{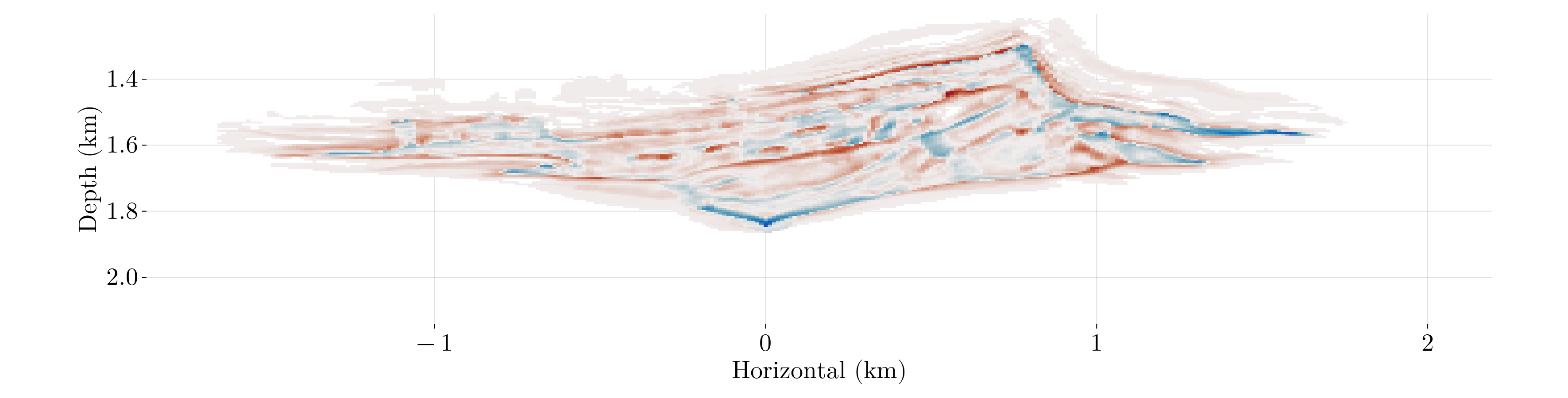}
    \caption{\enkf/ analysis error}
\end{subfigure}
\caption{\CO2 saturation error after 5 years for NoObs and \enkf/.}
\label{fig:alg-plume-errors}
\end{figure*}

\Cref{fig:alg-plumes} shows a comparison of the final plumes for \enkf/ and NoObs, and \Cref{fig:alg-plume-errors} shows the error in the final plumes.
While the ground-truth plume has sharp edges due to the sharp edges in the true permeability field, the estimates are smoother due to each ensemble member having different realizations of the permeability field with different locations for the sharp edges.
Furthermore, in order of from most smooth to least smooth, we have the NoObs plume, the \enkf/ prediction, the \enkf/ analysis, and the ground-truth.
This shows that assimilating observations achieves sharper plume estimates, both in forecasting (the \enkf/ prediction) and filtering (the \enkf/ analysis).

The most notable differences occur in areas of low permeability in the ground-truth.
Specifically, the ground-truth has empty pockets that the \CO2 does not reach due to the low permeability, and the ground-truth \CO2 saturation does not spread out as much.
The NoObs estimate misses all except the single pocket at 1.5 km depth, 0.6 km horizontal.
The \enkf/ recovers that pocket and shows signs of the 5 pockets along 1.6 km depth from -0.3 km to 0.3 km horizontal.
We note it is very hard to recover these without updating the ensemble members' permeabilities, as each time the transition operator is applied, \CO2 flows into the pockets for members that have high permeability there.

Furthermore, the estimated states show larger plume boundaries due to the uncertain permeability allowing the \CO2 to spread faster for some permeability samples. The \enkf/ estimate shows signs of correcting for this, seen by the noisy scatterings of minuscule but nonzero saturations on the edge of the estimate plume.
This is due to the \enkf/ updating the saturation based on seismic data.
Saturation updates outside the boundary of the ground-truth plume tend to push the saturation to be small or even negative, which is then clamped to zero.

\begin{figure}[htbp]
\centering
\begin{subfigure}{0.45\textwidth}
    \centering
    \includegraphics[width=\textwidth]{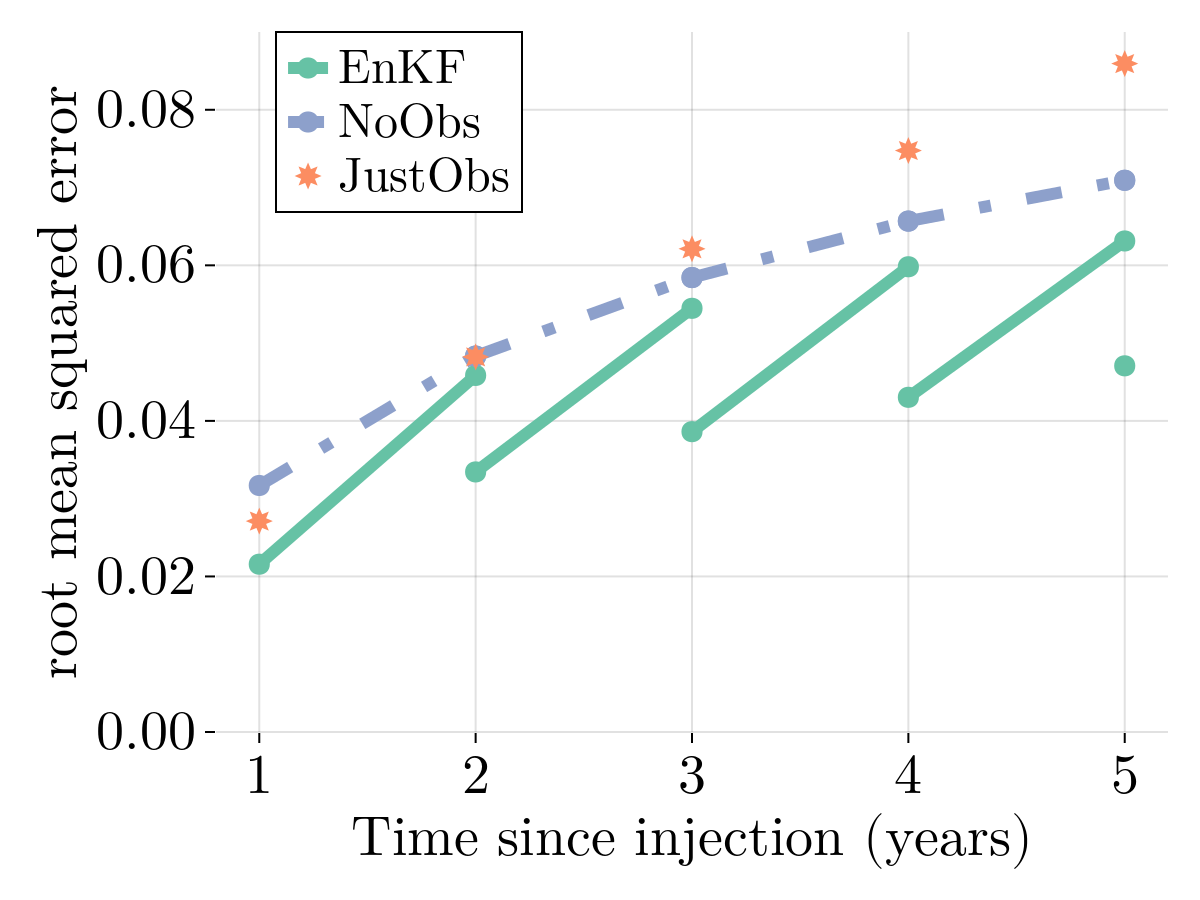}
    \label{fig:l2-errors}
\end{subfigure}
\begin{subfigure}{0.45\textwidth}
    \centering
    \includegraphics[width=\textwidth]{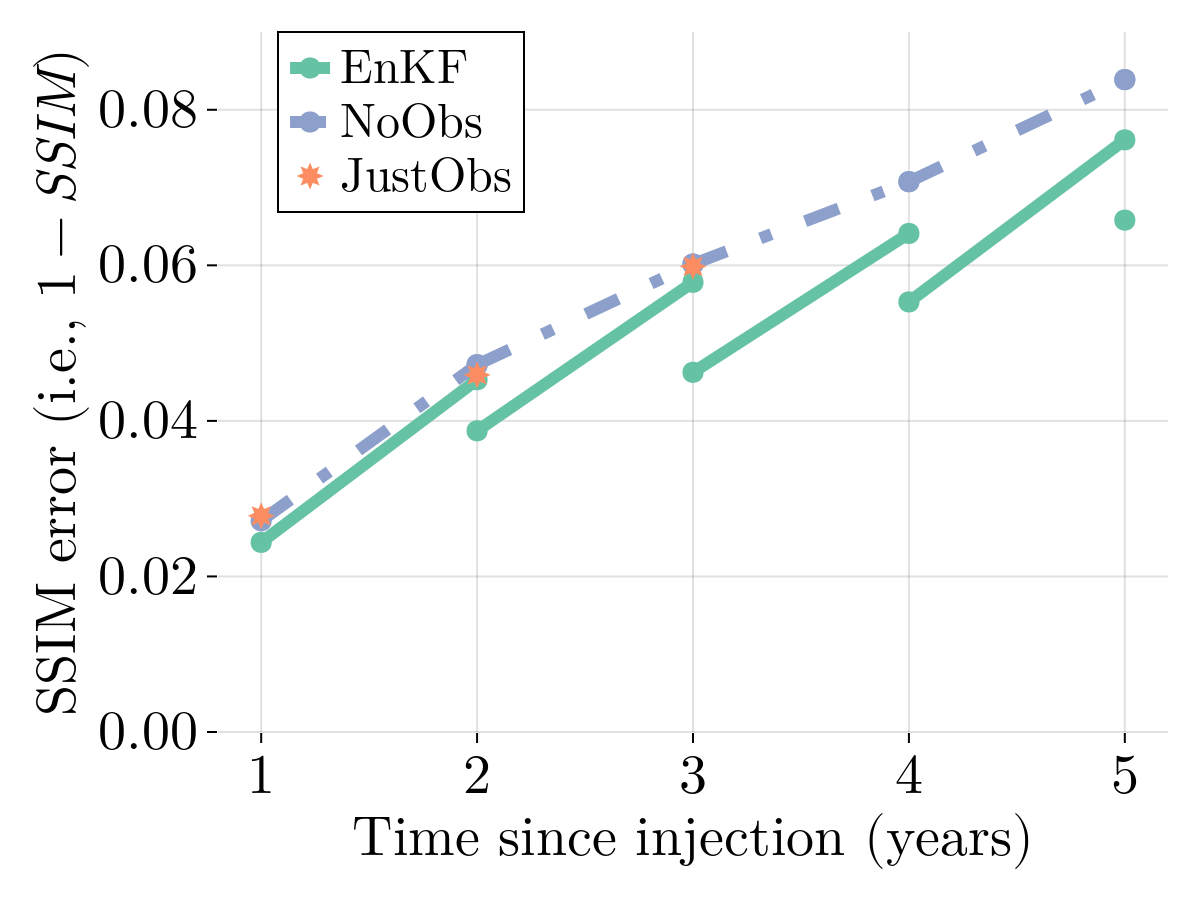}
    \label{fig:ssim-errors}
\end{subfigure}
\caption{
Error in predicted \CO2 saturation over time.
The error between observation steps are linearly interpolated.
These JustObs errors were generated with non-noisy observations and a hybrid $\ell_1$/$\ell_2$ regularization, which achieved the lowest error out of the norms we applied.
The SSIM error for JustObs is too high to be shown at years 4 and 5.
}
\label{fig:filter-scalar-errors}
\end{figure}

\Cref{fig:filter-scalar-errors} shows the error for each method over time, with discontinuities when observations are assimilated.
The x-axis starts at year 1 when the first observation is assimilated, because, before observations, each method is identical.
JustObs was unable to reduce the RMSE with noisy observations, so we show JustObs results solely for non-noisy data.
The NoObs error growth over time shows how much the ensemble's distribution of permeability fields causes the mean estimate to diverge from the ground-truth plume.
Comparing the JustObs error with the typically lower NoObs error, we conclude the permeability's uncertainty in this experiment is small enough that the knowledge of the \CO2 dynamics is more informative than seismic measurements. Real applications may have much higher uncertainty in the \CO2 dynamics, which may make JustObs perform better than NoObs.

The \enkf/ combines the knowledge of the \CO2 dynamics with the seismic measurements and consistently achieves lower error than NoObs and JustObs, thereby showing the benefit of data assimilation.
The uncertainty in the permeability causes a sharp increase in error between observations, which shows a limited ability for making accurate predictions.
Future research can use the seismic observations to reduce the uncertainty in the \CO2 dynamics, which will decrease the growth in error between observations.

\subsection{\enkf/ noise parameter tests}

\begin{figure}[htbp]
\centering
\begin{subfigure}{0.45\textwidth}
    \centering
    \includegraphics[width=\textwidth]{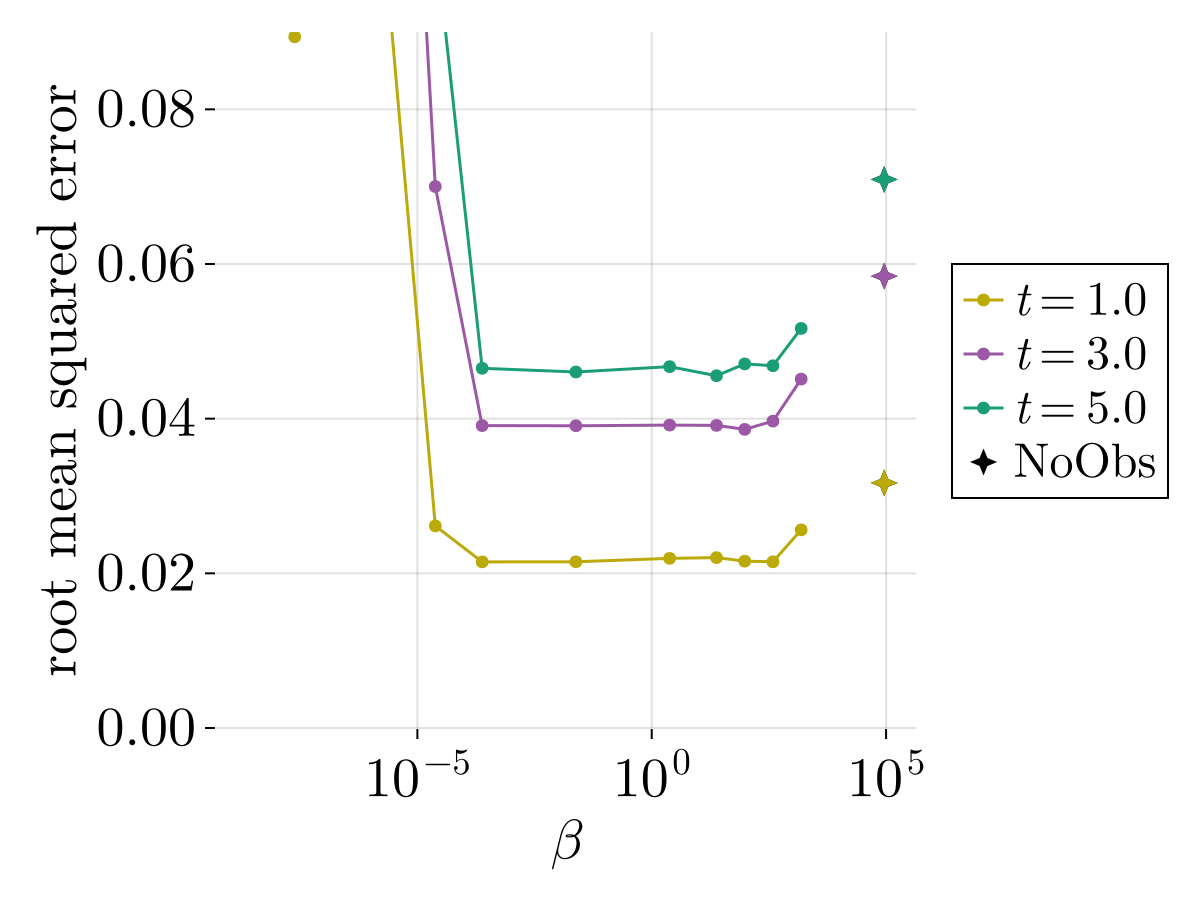}
    \caption{$\alpha = 1$}
    \label{fig:alpha1-beta-tests}
\end{subfigure}
\begin{subfigure}{0.45\textwidth}
    \centering
    \includegraphics[width=\textwidth]{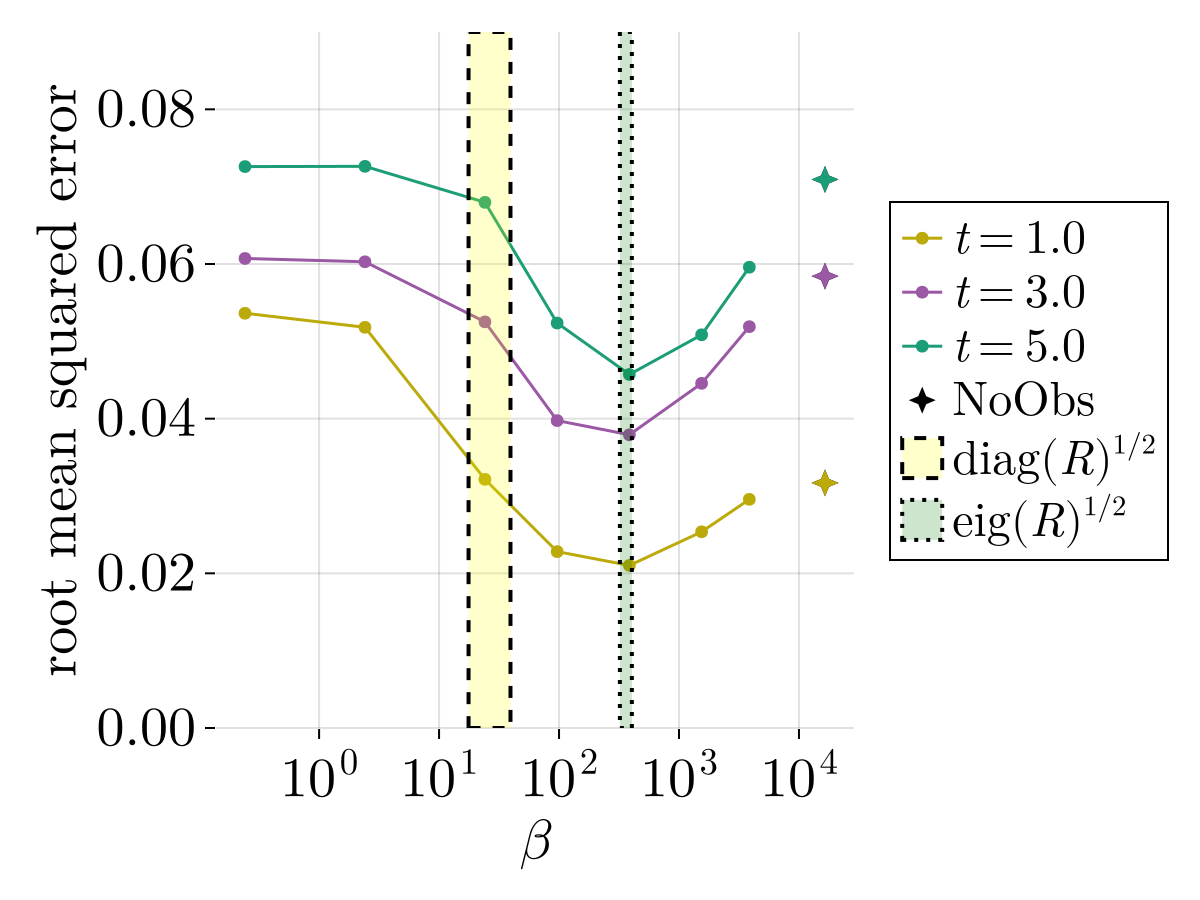}
    \caption{$\alpha = 0$}
    \label{fig:alpha0-beta-tests}
\end{subfigure}
\caption{RMSE vs $\beta$ ($\text{MPa}^2 \cdot \text{km} / \text{MRayl}$) with 8 dB SNR ($\nu = \nu^* \approx 0.4$) at three time steps after the \enkf/ update.
In \subref{fig:alpha1-beta-tests}, the noise covariance includes simulated noise plus a diagonal identity scaled by $\beta$. A wide range of $\beta$ achieves similar error, with a sharp increase in error for small regularization.
In \subref{fig:alpha0-beta-tests}, the noise covariance is approximated solely as a diagonal matrix, causing more sensitivity to the choice of $\beta$.
In either case, the error with large regularization should approach the NoObs case, which is equivalent to $\beta = \infty$.
}
\label{fig:beta-tests}
\end{figure}

\begin{figure}[htbp]
\centering
\includegraphics[width=0.45\textwidth]{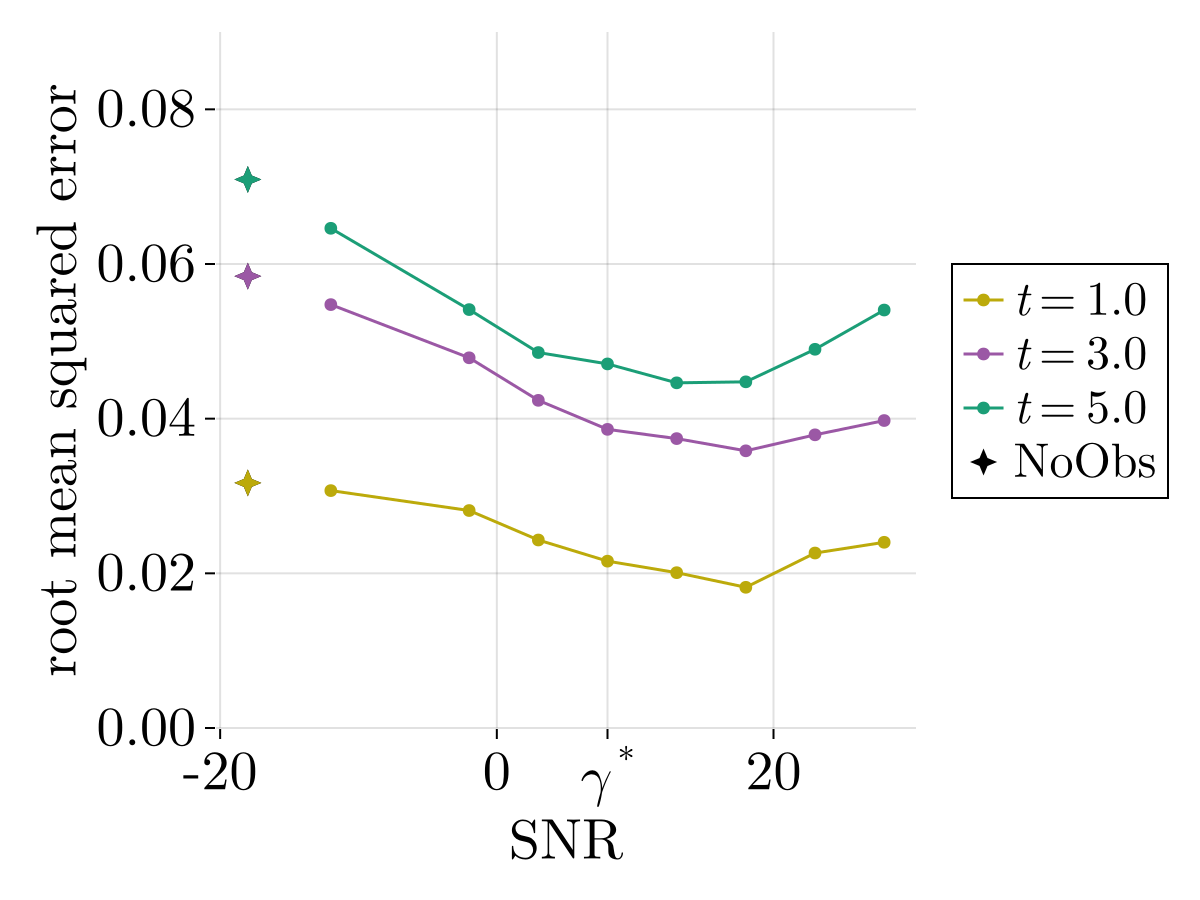}
\caption{RMSE vs simulated SNR $\gamma = -20 \log_{10} \nu$ with fixed ground-truth SNR $\gamma^* = 8$ dB, $\beta = 96.5 \;\text{MPa}^2 \cdot \text{km} / \text{MRayl}$, and $\alpha = 1$ at three time steps after the \enkf/ update.
Underestimating the SNR should approach the NoObs case ($\text{SNR} = -\infty$) as observations are treated as uninformative.
Overestimating the SNR leads to an increase in error because the filter starts to fit the noise.
}
\label{fig:nu-tests}
\end{figure}

\begin{figure}[htbp]
\centering
\includegraphics[width=0.45\textwidth]{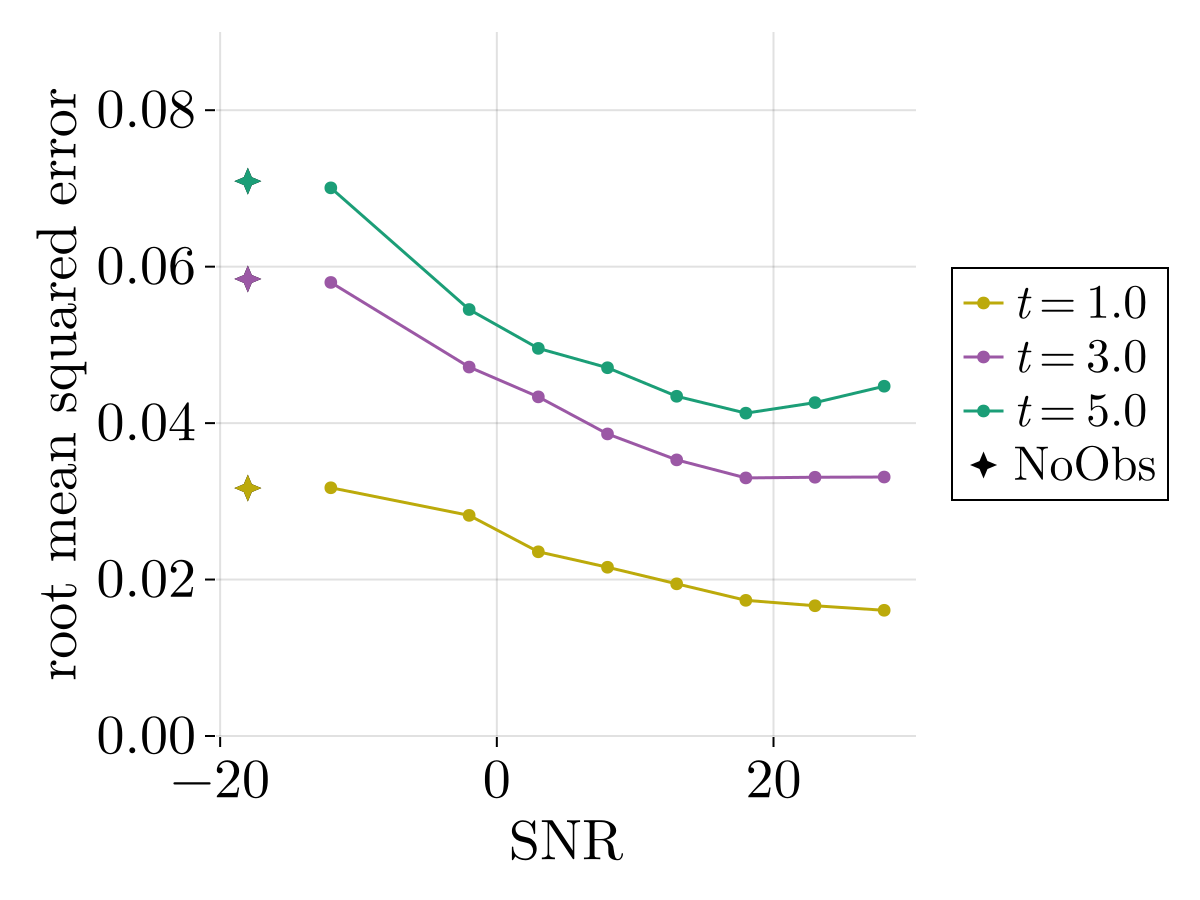}
\caption{RMSE vs ground-truth SNR $\gamma^* = -20 \log_{10} \nu^*$ with $\nu = \nu^*$ and fixed  $\beta = 96.5 \;\text{MPa}^2 \cdot \text{km} / \text{MRayl}$ and $\alpha = 1$ at three time steps after the \enkf/ update.
Low SNR (large noise) should approach the NoObs case ($\text{SNR} = -\infty$) in the two-norm (in expectation).
The post-assimilated error increases with SNR at later time steps.
That may indicate $\nu^*$ does not capture all the sources of noise, meaning there may be noise even when $\nu^* = 0$  due to numerical simulation.
To avoid this issue, it may be better to overestimate the noise instead of underestimating it.}
\label{fig:nutrue-tests}
\end{figure}

By comparing to the baselines, we conclude that the \enkf/ with our initial choice of \enkf/ algorithm parameters performs well.
Now, we examine performance when modifying the \enkf/ parameters.
\Cref{fig:beta-tests,fig:nu-tests,fig:nutrue-tests} show the results, discussed below, of testing values for the different noise parameters in \cref{eq:noisy_tests_enkf_update}, duplicated here,
\begin{multline}
\recallLabel[revisited]{eq:noisy_tests_enkf_update}.
\end{multline}

\paragraph*{Regularization tests}
We first vary $\beta$ with $\nu = \nu^*$ to ensure the \enkf/ performance is not sensitive to the magnitude of regularization.
$\beta$ scales the regularization covariance $R=\nu^2 \beta^2 I$ in the observation covariance matrix, and has units of $\text{MPa}^2 \cdot \text{km} / \text{MRayl}$.
If $\alpha = 0$, then $R$ should be the identity scaled by the variance of the noise.
By sampling the noise, we form an estimate of the true noise covariance and empirically determine the average noise standard deviation across the seismic observations to be $\nu \beta \approx 11$ for an SNR of 8 dB, which corresponds to $\nu = 10^{-8 \text{ dB} /20} \approx 0.4$ and $\beta \approx 28$.
Scaling by the average noise standard deviation is the typical method for regularizing the observation covariance inversion in the \enkf/.
We additionally compute the largest 256 eigenvalues of the sample estimate of the true noise covariance, as the largest eigenvalue leads to a more unbiased estimate of the regularized observation noise covariance when approximating correlated noise with a multiple of the identity. This leads to an estimate of $\beta \approx 400$.

In \cref{fig:alpha1-beta-tests}, we plot the error as a function of $\beta$ with simulated noise in the observation covariance ($\alpha = 1$) and determine that a wide range of $\beta$ values from $10^{-4}$ to $10^{3}$ achieve equivalent results.
The error sharply increases for very small $\beta$ approximately six orders of magnitude smaller than the true noise standard deviation.
As $\beta$ becomes very large, the post-assimilation error begins to approach the pre-assimilation error as expected.
We conclude that if the noise is included in the observation covariance ($\alpha = 1$), then the regularization should be chosen within a couple of orders of magnitude of the true noise standard deviation and err on the side of being smaller.

We then vary $\beta$ without simulated noise in the observation covariance ($\alpha = 0$) to see if simulating the noise gives the \enkf/ better performance than fully approximating the noise covariance as a multiple of the identity matrix.
In \cref{fig:alpha0-beta-tests}, we plot the error as a function of $\beta$.
We find that for $\alpha = 0$, the error is much more sensitive to the magnitude of the regularization.
Furthermore, we highlight the range of $\beta$ based on the diagonal of the observation noise sample covariance and the range of $\beta$ based on the largest eigenvalues.
We conclude the minimum error is achieved when choosing the regularization magnitude based on the largest eigenvalues.
This can be useful for systems where simulating noise is more costly.
For example, the regularization estimate can be computed offline by computing noise samples and the resulting noise covariance eigenvalues.

For full waveform observations, simulating noise is inexpensive compared to the cost of the transition and observation operators, and the robustness in terms of the choice of $\beta$ parameter when $\alpha=1$ offsets the low cost of simulating the noise.

\paragraph*{Simulated noise tests}
We then vary $\nu$ with $\nu^*$ fixed and simulating the observation noise samples ($\alpha = 1$) to ensure the \enkf/ performance is not sensitive to the estimate of the observation noise magnitude.
In \cref{fig:nu-tests}, we plot the error versus the simulated SNR in decibels, which is $-20 \log_{10} \nu$.
As expected, when the estimated noise magnitude is chosen very large (low SNR), the filter approaches the NoObs results, essentially ignoring observations. 
On the opposite end, when the simulated SNR is too high, the filter puts too much weight on the observations, thus overfitting them and raising the error.

Estimating the noise correctly with estimated SNR = $\gamma^*$ should give the lowest error in the long term.
However, in the short term, a large error in the initial prior can make it advantageous to weight observations higher in earlier time steps.
In \cref{fig:nu-tests} at year 1, the minimum RMSE is achieved by overestimating the SNR by 10 dB, but by year 5, the minimum RMSE is achieved by overestimating SNR by 5 dB.
As more observations are collected, the effect of the initial prior decreases, and the minimum should shift to coincide with the true SNR $\gamma^*$.

\paragraph*{True noise tests}
Finally, we vary $\nu^*$ with $\nu = \nu^*$ and simulating the observation noise samples ($\alpha = 1$) to ensure the \enkf/ can still perform well on varying levels of noisy data.
In \cref{fig:nutrue-tests}, we plot the error versus the true SNR in decibels.
As expected, we find that the \enkf/ error approaches the NoObs error for large noise magnitudes (low SNR).
Except for the first two time steps, we find that the error increases as the noise becomes very small.
This can be explained by the results with large SNR in the $\nu$ tests in \cref{fig:nu-tests} that have large error due to overfitting the noise.
As we decrease $\nu^*$ and $\nu$ together, we see better results, until we reach a point where the $\nu = \nu^*$ does not properly account for the noise. Specifically, for high SNR, we expect simulation error to dominate the noise, and therefore, the error increases.
This is not a problem in real scenarios as the real-world seismic noise always dominates floating-point arithmetic errors.

\section{Conclusion}

The existing literature has shown various Kalman filters applied to monitoring \CO2 plume on a relatively small scale or without seismic measurements or without \CO2 dynamics.
We apply the \enkf/ to a high-dimensional \CO2 plume monitored with seismic measurements.
We show that it achieves lower error than the non-data-assimilation baselines, which ignore the \CO2 dynamics or ignore the observations.
The \enkf/ is thus a promising tool to use for monitoring \CO2 reservoirs.

In addition to comparing to the two baselines, we also examine the sensitivity of the \enkf/ to the choice of noise parameters.
Lower error is achieved by simulating the observation noise and implicitly obtaining a sample observation noise covariance, compared to approximating the observation noise covariance as a diagonal matrix.
Since simulating noise for seismic waveform measurements is inexpensive, we recommend doing this to increase the robustness with respect to the choice of regularization magnitude in the \enkf/ observation covariance.
If the noise is not simulated, we see that the regularization parameter should be chosen based on the largest eigenvalues, not the average standard deviation.
Regarding the magnitude of the simulated noise, we find slightly better performance with overestimating the true noise, but this result is specific to the case of relatively low uncertainty in the \CO2 dynamics and may not apply in the general case.

\paragraph*{Future work}
We considered all parameters to be known except for the permeability and the \CO2 saturation and pressure, and we updated only the \CO2 saturation when assimilating observations.
Because we do not update the permeability, the error in the \enkf/ forecast states is very similar to the error from ignoring observations.
To achieve accurate predictions of the \CO2 plume, the geological parameters controlling the flow must be updated as well.
Therefore, we plan to apply the \enkf/ to a high-dimensional \CO2 plume with \emph{nonlinear} seismic observations \emph{and} update the permeability and porosity fields along with the \CO2 saturation, following the work of \textcite{li_smoothingbased_2017} who updated the permeability and \CO2 saturation with an \enkf/ variant,
\textcite{ma_dynamic_2019} who updated the permeability and porosity based on the \enkf/ with approximated seismic images, and \textcite{li_coupled_2020,yin_time-lapse_2024} who demonstrated end-to-end inversion for permeability based on time-lapse FWI of \CO2 plumes.

\makeatletter
\printbibliography
\makeatother

\end{document}

%% file: common.tex
\usepackage{arxiv}

\usepackage[urlcolor=blue,citecolor=black,linkcolor=black]{hyperref}

\usepackage[ruled,vlined,linesnumbered]{algorithm2e}%

\SetKwInOut{Input}{Input}
\DontPrintSemicolon
\SetAlgoLined
\SetKw{KwForIn}{in}
\SetKwData{Prior}{prior}
\SetKwData{Prediction}{prior}
\SetKwData{Posterior}{posterior}
\SetKwData{State}{p\textunderscore x}
\SetKwData{NewData}{observations}
\SetKwFunction{Advance}{Predict}
\SetKwFunction{Assimilate}{Update}
\SetKwProg{Fn}{Function}{:}{}
\SetKwComment{Comment}{// }{}

\DeclareUnicodeCharacter{2082}{\textunderscore{2}}
\DeclareUnicodeCharacter{2113}{$\ell$}

\usepackage{graphicx}
\usepackage{xcolor}
\usepackage{amsmath}
\usepackage{amssymb}
\usepackage{textalpha}%

\usepackage{subcaption}
\usepackage[mathlines]{lineno}%
\usepackage{amsfonts}%
\usepackage{etoolbox}%
\newcommand*\linenomathpatch[1]{%
  \cspreto{#1}{\linenomath}%
  \cspreto{#1*}{\linenomath}%
  \csappto{end#1}{\endlinenomath}%
  \csappto{end#1*}{\endlinenomath}%
}%
\newcommand*\linenomathpatchAMS[1]{%
  \cspreto{#1}{\linenomathAMS}%
  \cspreto{#1*}{\linenomathAMS}%
  \csappto{end#1}{\endlinenomath}%
  \csappto{end#1*}{\endlinenomath}%
}%
\expandafter\ifx\linenomath\linenomathWithnumbers%
  \let\linenomathAMS\linenomathWithnumbers%
  \patchcmd\linenomathAMS{\advance\postdisplaypenalty\linenopenalty}{}{}{}%
\else%
  \let\linenomathAMS\linenomathNonumbers%
\fi%
\linenomathpatch{equation}%
\linenomathpatchAMS{gather}%
\linenomathpatchAMS{multline}%
\linenomathpatchAMS{align}%
\linenomathpatchAMS{alignat}%
\linenomathpatchAMS{flalign}%
\newcommand\labelAndRemember[2]%
  {\expandafter\gdef\csname labeled:#1\endcsname{#2}%
   \label{#1}#2}%
\newcommand\recallLabel[2][]%
   {\csname labeled:#2\endcsname\tag{\ref{#2} #1}}%
\makeatletter%
\patchcmd{\mmeasure@}{\measuring@true}{%
  \measuring@true%
  \ifnum-\linenopenaltypar>\interdisplaylinepenalty%
    \advance\interdisplaylinepenalty-\linenopenalty%
  \fi%
  }{}{}%
\makeatother%

\def\CO2{CO\texorpdfstring{\textsubscript{2}}{\texttwoinferior}}
\def\Co2{\CO2}
\def\enkf/{EnKF}
\def\fdvar/{4D-Var}
\newcommand\defacronym[2]{\emph{#1} (#2)}

\newcommand\xs{N_x}
\newcommand\ys{N_y}
\newcommand\es{N_e}
\renewcommand\vec[1]{\mathbf{#1}}

\DeclareMathOperator{\cov}{cov}

\DeclareMathOperator{\RMSE}{RMSE}
\DeclareMathOperator*{\Expect}{\mathbb{E}}

\DeclareMathOperator*{\argmin}{arg\,min}

\usepackage[
  backend=biber,
  style=ieee,
  minnames=1,
  maxcitenames=2, maxbibnames=6
]{biblatex}

\DeclareSourcemap{
  \maps[datatype=bibtex, overwrite]{
    \map{
        \step[fieldset=address, null]
        \step[fieldset=location, null]
        \step[fieldset=issn, null]
        \step[fieldset=url, null]
        \step[fieldset=shorttitle, null]
        \step[fieldset=abstract, null]
        \step[fieldset=urldate, null]
        \step[fieldset=langid, null]
        \step[fieldset=keywords, null]
    }
  }
}

\author{
    Grant Bruer \\
    School of Computational Science and Engineering \\
    Georgia Institute of Technology \\
    Atlanta, GA 30332 USA \\
    \And
    Abhinav Prakash Gahlot \\
    School of Earth and Atmospheric Sciences \\
    Georgia Institute of Technology \\
    Atlanta, GA 30332 USA \\
    \And
    Edmond Chow \\
    School of Computational Science and Engineering \\
    Georgia Institute of Technology \\
    Atlanta, GA 30332 USA \\
    \And
    Felix Herrmann \\
    School of Computational Science and Engineering \\
    Georgia Institute of Technology \\
    Atlanta, GA 30332 USA \\
}

\usepackage[capitalise,noabbrev]{cleveref}
\newdimen\imageheight